\documentclass[10pt]{acmconfbig}
\usepackage{amssymb}
\usepackage{amsmath}
\usepackage{prooftree}
\newtheorem{lemma}{Lemma}
\newtheorem{theorem}{Theorem}
\newtheorem{definition}{Definition}
\newenvironment{proof}{\textit{Proof:}}{}
\newcommand{\qed}{\hfill$\Box$}
\newcommand{\mktype}[1]{\mathsf{#1}}
\newcommand{\Int}{\mktype{Int}}
\newcommand{\Unit}{\mktype{Unit}}
\newcommand{\Qbit}{\mktype{Qbit}}
\newcommand{\Bit}{\mktype{Bit}}
\newcommand{\Op}[1]{\mktype{Op}(#1)}
\newcommand{\Chant}[1]{\widehat{~}\,[#1]}
\newcommand{\Listt}[1]{{#1}~\mktype{List}}
\newcommand{\Prodt}[2]{{#1}*{#2}}
\newcommand{\kw}[1]{\mathsf{#1}}
\newcommand{\mkterm}[1]{\mathsf{#1}}
\newcommand{\num}[1]{\mkterm{#1}}
\newcommand{\msure}[1]{\mkterm{measure}~{#1}}
\newcommand{\plus}[2]{{#1}\mkterm{+}{#2}}
\newcommand{\unit}{\mkterm{unit}}
\newcommand{\ctxt}[2]{{#1}[#2]}
\newcommand{\bnf}{::=}
\newcommand{\alt}{~|~}
\newcommand{\prob}[2]{{#1}\bullet{#2}}
\newcommand{\bp}{\boxplus}
\newcommand{\Prob}[1]{\boxplus_{#1}}
\newcommand{\ms}[1]{|{#1}|^{2}}
\newcommand{\cnfig}[3]{({#1};{#2};{#3})}

\newcommand{\ptrns}[3]{{#1}\stackrel{#2}{\longrightarrow}{#3}}
\newcommand{\vred}[2]{{#1}\longrightarrow_{\mathsf{v}}{#2}}
\newcommand{\ered}[2]{{#1}\longrightarrow_{\mathsf{e}}{#2}}
\newcommand{\cred}[2]{{#1}\longrightarrow{#2}}
\newcommand{\subst}[3]{{#1}\{{#2}/{#3}\}}

\newcommand{\mkRrule}[1]{\mbox{\textsc{R-#1}}}
\newcommand{\mkSrule}[1]{\mbox{\textsc{S-#1}}}
\newcommand{\mkTrule}[1]{\mbox{\textsc{T-#1}}}
\newcommand{\mkITrule}[1]{\mbox{\textsc{IT-#1}}}
\newcommand{\Rplus}{\mkRrule{Plus}}
\newcommand{\Rmeasure}{\mkRrule{Measure}}
\newcommand{\Rcontext}{\mkRrule{Context}}
\newcommand{\Rtrans}{\mkRrule{Trans}}
\newcommand{\Rexpr}{\mkRrule{Expr}}
\newcommand{\Rcom}{\mkRrule{Com}}
\newcommand{\Rnew}{\mkRrule{New}}
\newcommand{\Rqbit}{\mkRrule{Qbit}}
\newcommand{\Rpar}{\mkRrule{Par}}
\newcommand{\Rcong}{\mkRrule{Cong}}
\newcommand{\Ract}{\mkRrule{Act}}
\newcommand{\Rprob}{\mkRrule{Prob}}
\newcommand{\Rperm}{\mkRrule{Perm}}
\newcommand{\Tintlit}{\mkTrule{IntLit}}
\newcommand{\Tunit}{\mkTrule{Unit}}
\newcommand{\Top}{\mkTrule{Op}}
\newcommand{\Tvar}{\mkTrule{Var}}
\newcommand{\Tmeasure}{\mkTrule{Msure}}
\newcommand{\Ttrans}{\mkTrule{Trans}}
\newcommand{\Tnil}{\mkTrule{Nil}}
\newcommand{\Tpar}{\mkTrule{Par}}
\newcommand{\Tin}{\mkTrule{In}}
\newcommand{\Tout}{\mkTrule{Out}}
\newcommand{\Taction}{\mkTrule{Act}}
\newcommand{\Tnew}{\mkTrule{New}}
\newcommand{\Tqbit}{\mkTrule{Qbit}}
\newcommand{\Tplus}{\mkTrule{Plus}}

\newcommand{\ITintlit}{\mkITrule{IntLit}}
\newcommand{\ITunit}{\mkITrule{Unit}}
\newcommand{\ITop}{\mkITrule{Op}}
\newcommand{\ITvar}{\mkITrule{Var}}
\newcommand{\ITmeasure}{\mkITrule{Msure}}
\newcommand{\ITtrans}{\mkITrule{Trans}}
\newcommand{\ITnil}{\mkITrule{Nil}}
\newcommand{\ITpar}{\mkITrule{Par}}
\newcommand{\ITin}{\mkITrule{In}}
\newcommand{\ITout}{\mkITrule{Out}}
\newcommand{\ITaction}{\mkITrule{Act}}
\newcommand{\ITnew}{\mkITrule{New}}
\newcommand{\ITqbit}{\mkITrule{Qbit}}
\newcommand{\ITplus}{\mkITrule{Plus}}
\newcommand{\ITidQ}{\mkITrule{IdQ}}
\newcommand{\ITidC}{\mkITrule{IdC}}
\newcommand{\Snil}{\mkSrule{Nil}}
\newcommand{\Scomm}{\mkSrule{Comm}}
\newcommand{\Sassoc}{\mkSrule{Assoc}}

\newcommand{\typed}[3]{{#1}\vdash{#2}:{#3}}
\newcommand{\ptyped}[2]{{#1}\vdash{#2}}
\newcommand{\ityped}[5]{{#1};{#2};{#3}\vdash{#4}:{#5}}
\newcommand{\iptyped}[4]{{#1};{#2};{#3}\vdash{#4}}

\newcommand{\length}[1]{|{#1}|}
\newcommand{\scong}{\equiv}
\newcommand{\ket}[1]{|#1\rangle}
\newcommand{\nil}{\mathbf{0}}
\renewcommand{\parallel}{\mathbin{\mid}}
\newcommand{\inp}[2]{{#1}?{[#2]}}
\newcommand{\outp}[2]{{#1}!{[#2]}}

\renewcommand{\vec}[1]{\widetilde{#1}}
\newcommand{\new}{\mathsf{new}\ }
\newcommand{\qbit}{\mathsf{qbit}\ }

\newcommand{\chant}[1]{\widehat{~}[#1]}
\newcommand{\tid}[2]{{#1}\mathrel{\!:\!}{#2}}
\newcommand{\ranget}[2]{\mathsf{{#1}..{#2}}}
\newcommand{\qgate}[1]{\mathsf{#1}}
\newcommand{\pname}[1]{\mathit{#1}}
\newcommand{\action}[1]{\{#1\}}
\newcommand{\trans}[2]{{#1}\mathbin{*\!\!=}{#2}}
\newcommand{\sep}{\,.\,}

\newcommand{\measure}{\mathsf{measure}\ }

\newcommand{\dom}[1]{\mathit{dom}(#1)}
\newcommand{\fc}[1]{\mathit{fc}(#1)}
\newcommand{\fv}[1]{\mathit{fv}(#1)}
\newcommand{\fq}[1]{\mathit{fq}(#1)}

\newcommand{\vect}[2]{\begin{pmatrix}{#1}\\{#2}\end{pmatrix}}
\newcommand{\matr}[4]{\begin{pmatrix}{#1}&{#2}\\{#3}&{#4}\end{pmatrix}}

\title{Communicating Quantum Processes}
\author{
\begin{tabular}{c@{\extracolsep{20mm}}c}
Simon J.\ Gay & Rajagopal Nagarajan\thanks{Partially supported by
the UK EPSRC (GR/S34090) and the EU Sixth Framework Programme (Project SecoQC).} \\
Department of Computing Science & Department of Computer Science \\
University of Glasgow, UK & University of Warwick, UK \\
\texttt{simon@dcs.gla.ac.uk} & \texttt{biju@dcs.warwick.ac.uk} \\ \\
\multicolumn{2}{c}{July 16, 2004}
\end{tabular}
}
\begin{document}
\maketitle

\begin{abstract}
We define a language CQP (Communicating Quantum Processes) for
modelling systems which combine quantum and classical communication
and computation. CQP combines the communication primitives of the
pi-calculus with primitives for measurement and transformation of
quantum state; in particular, quantum bits (qubits) can be transmitted
from process to process along communication channels. CQP has a static
type system which classifies channels, distinguishes between quantum
and classical data, and controls the use of quantum state. We formally
define the syntax, operational semantics and type system of CQP, prove
that the semantics preserves typing, and prove that typing guarantees
that each qubit is owned by a unique process within a system. We
illustrate CQP by defining models of several quantum communication
systems, and outline our plans for using CQP as the foundation for
formal analysis and verification of combined quantum and classical systems.
\end{abstract}

\toappear{An earlier version of this paper is in \emph{Proceedings of
the 2nd International Workshop on Quantum Programming Languages},
Turku Centre for Computer Science General Publication No.\ 33, June 2004.}

\section{Introduction}
\label{sec-intro}
Quantum computing and quantum communication have attracted growing
interest since their inception as research areas more than twenty years
ago, and there has been a surge of activity among computer scientists
during the last few years. While quantum computing offers the prospect
of vast improvements in algorithmic efficiency for certain problems,
quantum cryptography can provide communication systems which
will be secure even in the presence of hypothetical future quantum computers.
As a practical technology, quantum communication has progressed far
more rapidly than quantum computing. Secure communication involving
quantum cryptography has recently been demonstrated in a scenario
involving banking transactions in Vienna
\cite{PoppeA:praqkd}, systems are commercially available from
Id Quantique, MagiQ Technologies and NEC, and plans
have been reported to establish a nationwide quantum communication
network in Singapore. Secure quantum communication will undoubtedly become a
fundamental part of the technological infrastructure of society,
long before quantum computers can tackle computations of a useful size.

However, secure quantum communication is not a solved
problem. Although particular protocols have been mathematically proved
correct (for example, Mayers' analysis
\cite{MayersD:uncsqc} of the Bennett-Brassard protocol (BB84)
\cite{BennettCH:quacpd} for quantum key distribution), this does not
guarantee the security of systems which use them. Experience of
classical security analysis has shown that even if \emph{protocols}
are theoretically secure, it is difficult to achieve robust and
reliable implementations of secure \emph{systems}: security can be
compromised by flaws at the implementation level or at the boundaries
between systems. To address this problem, computer scientists have
developed an impressive armoury of techniques and tools for formal
modelling, analysis and verification of classical security protocols
and communication systems which use them
\cite{RyanP:modasp}. These techniques have been remarkably successful both in
establishing the security of new protocols and in demonstrating flaws
in protocols which had previously been believed to be secure. Their
strength lies in the ability to model \emph{systems} as well as idealized
protocols, and the flexibility to easily re-analyze variations in
design.

Our research programme is to develop techniques and tools for formal
modelling, analysis and verification of quantum communication and
cryptographic systems. More precisely we aim to handle systems which
combine quantum and classical communication and computation, for two
reasons: the first quantum communication systems will implement
communication between classical computers; and protocols such as BB84
typically contain classical communication and computation as well as
quantum cryptography. We cannot simply make use of existing techniques
for classical security analysis: for example, treating the security of
quantum cryptography axiomatically would not permit analysis of the
protocols which \emph{construct} quantum cryptographic
keys. Furthermore, the inherently probabilistic nature of quantum
systems means that not all verification consists of checking absolute
properties; we need a probabilistic modelling and analysis framework.

Any formal analysis which involves automated tools requires a
modelling language with a precisely-defined semantics. The purpose of
this paper is to define a language, CQP (Communicating Quantum
Processes), which will serve as the foundation for the programme
described above. CQP combines the communication primitives of the
pi-calculus \cite{MilnerR:calmpfull,SangiorgiD:pictm} with primitives
for transformation and measurement of quantum state. In particular,
qubits (quantum bits, the basic elements of quantum data) can be
transmitted along communication channels. In
Section~\ref{sec-examples} we introduce CQP through a series of
examples which cover a wide spectrum of quantum information processing
scenarios: a quantum coin-flipping game; a quantum communication
protocol known as teleportation; and a quantum
bit-commitment protocol. The latter will lead naturally to a model of the
BB84 quantum key-distribution protocol in future work. In
Section~\ref{sec-syntax} we formalize the syntax of CQP and define an
operational semantics which combines non-determinism (arising in the
same way as in pi-calculus) with the probabilistic results of quantum
measurements. In Section~\ref{sec-types} we define a static type
system which classifies data and communication channels, and crucially
treats qubits as physical resources: if process $P$ sends qubit $q$ to
process $Q$, then $P$ must not access $q$ subsequently, and this
restriction can be enforced by static typechecking. In
Section~\ref{sec-soundness} we prove that the invariants of the type
system are preserved by the operational semantics, guaranteeing in
particular that at every point during execution of a system, every qubit is
uniquely owned by a single parallel component. In
Section~\ref{sec-future} we outline our plans for further work,
focusing on the use of both standard (non-deterministic) and
probabilistic model-checking systems.

\subsection*{Related Work}
There has been a great deal of interest in quantum programming languages,
resulting in a number of proposals in different styles, for example
\cite{KnillE:conqp,OmerB:quapq,SandersJW:quap,SelingerP:towqpl,vanTonderA:lamcqc}.
Such languages can express arbitrary quantum state transformations and
could be used to model quantum protocols in those terms. However, our
view is that any model lacking an explicit treatment of communication
is essentially incomplete for the analysis of protocols; certainly in
the classical world, standard programming languages are not considered
adequate frameworks in which to analyze or verify
protocols. Nevertheless, Selinger's functional language QPL
\cite{SelingerP:towqpl} in particular has influenced our choice of
computational operators for CQP.

The closest work to our own, developed simultaneously but
independently, is Jorrand and Lalire's QPAlg \cite{JorrandP:towqpa}, which
also combines process-calculus-style communication with
transformation and measurement of quantum state. The most distinctive
features of our work are the type system and associated proofs, the
explicit formulation of an expression language which can easily be
extended, and our emphasis on a methodology for formal verification.

The work of Abramsky and Coecke
\cite{AbramskyS:catsqp} is also relevant. They define a
category-theoretic semantic foundation for quantum protocols, which
supports reasoning about systems and exposes deep connections between
quantum systems and programming language semantics, but they do not
define a formal syntax in which to specify models.  It will be
interesting to investigate the relationship between CQP and the
semantic structures which they propose.

\section{Preliminaries}
\label{sec-prelim}
We briefly introduce the aspects of quantum theory which are needed
for the rest of the paper. For more detailed presentations we refer
the reader to the books by Gruska \cite{GruskaJ:quac} and Nielsen and
Chuang \cite{NielsenMA:quacqi}. Rieffel and Polak
\cite{RieffelEG:intqcn} give 
an account aimed at computer scientists.

A \emph{quantum bit} or \emph{qubit} is a physical system which has
two states, conventionally written $\ket{0}$ and $\ket{1}$,
corresponding to one-bit classical values. These could be, for
example, spin states of a particle or polarization states of a photon,
but we do not consider physical details. According to quantum theory,
a general state of a quantum system is a \emph{superposition} or
linear combination of basis states. Concretely, a qubit has state
$\alpha\ket{0}+\beta\ket{1}$, where $\alpha$ and $\beta$ are complex
numbers such that $\ms{\alpha}+\ms{\beta}=1$; states which differ only
by a (complex) scalar factor with modulus $1$ are
indistinguishable. States can be represented by column vectors:
\[
\vect{\alpha}{\beta} = \alpha\ket{0}+\beta\ket{1}.
\]
Superpositions are illustrated by the quantum
coin-flipping game which we discuss in Section~\ref{sec-coinflip}.
Formally, a quantum state is a unit vector in a Hilbert space, i.e.\ a
complex vector space equipped with an inner product satisfying certain
axioms. In this paper we will restrict attention to collections of qubits. 

The basis $\{\ket{0},\ket{1}\}$ is known as the \emph{standard}
basis. Other bases are sometimes of interest, especially the
\emph{diagonal} (or \emph{dual}, or \emph{Hadamard}) basis consisting
of the vectors $\ket{+} = \frac{1}{\sqrt{2}}(\ket{0}+\ket{1})$ and
$\ket{-} = \frac{1}{\sqrt{2}}(\ket{0}-\ket{1})$. For example, with respect to
the diagonal basis, $\ket{0}$ is in a superposition of basis states:
\[
\ket{0} = \frac{1}{\sqrt{2}}\ket{+} +
\frac{1}{\sqrt{2}}\ket{-}.
\]

Evolution of a closed quatum system can be described by a 
\emph{unitary transformation}. If the state of a qubit is represented
by a column vector then a unitary transformation $U$ can be
represented by a complex-valued matrix $(u_{ij})$ such that $U = U^*$,
where $U^*$ is the conjugate-transpose of $U$ (i.e.\ element $ij$ of
$U^*$ is $\bar{u}_{ji}$). $U$ acts by matrix multiplication:
\[
\vect{\alpha'}{\beta'} = \matr{u_{00}}{u_{01}}{u_{10}}{u_{11}}\vect{\alpha}{\beta}
\]
A unitary transformation can also be defined by its effect on basis
states, which is extended linearly to the whole space. For example,
the \emph{Hadamard} transformation is defined by
\[
\begin{array}{lcl}
\ket{0} & \mapsto & \frac{1}{\sqrt{2}}\ket{0}+\frac{1}{\sqrt{2}}\ket{1} \\
\ket{1} & \mapsto & \frac{1}{\sqrt{2}}\ket{0}-\frac{1}{\sqrt{2}}\ket{1}
\end{array}
\]
which corresponds to the matrix
\[
\qgate{H} = \frac{1}{\sqrt{2}}\matr{1}{1}{1}{-1}.
\]
The Hadamard transformation creates superpositions:
$\qgate{H}{\ket{0}} = \ket{+}$ and $\qgate{H}{\ket{1}} = \ket{-}$.
We will also make use of the \emph{Pauli} transformations, denoted by
either $I,\sigma_x,\sigma_y,\sigma_z$ or $\sigma_0,\sigma_1,\sigma_2,\sigma_3$:
\[
\begin{array}{@{\extracolsep{2mm}}cccc}
I / \sigma_0 & \sigma_x / \sigma_1 & \sigma_y / \sigma_2 & \sigma_z / \sigma_3 \\ \\
\matr{1}{0}{0}{1} & \matr{0}{1}{1}{0} &
\matr{0}{-i}{i}{0} & \matr{1}{0}{0}{-1}
\end{array}
\]

A key feature of quantum physics is the r\^{o}le of
\emph{measurement}. If a qubit is in the state
$\alpha\ket{0}+\beta\ket{1}$ then measuring its value gives the result
$0$ with probability $\ms{\alpha}$ (leaving it in state $\ket{0}$) and
the result $1$ with probability $\ms{\beta}$ (leaving it in state
$\ket{1}$). Protocols sometimes specify measurement with respect to a
different basis, such as the diagonal basis; this can be expressed as
a unitary change of basis followed by a measurement with respect to
the standard basis. Note that if a qubit is in state
$\ket{+}$ then a measurement with respect
to the standard basis give result $0$ (and state $\ket{0}$) with
probability $\frac{1}{2}$, and result $1$ (and state $\ket{1}$) with
probability $\frac{1}{2}$. If a qubit is in state $\ket{0}$ then a
measurement with respect to the diagonal basis gives
result\footnote{Strictly speaking, the outcome of the measurement is
just the final state; the specific association of numerical results with final
states is a matter of convention.} $0$ (and
state $\ket{+}$) with probability
$\frac{1}{2}$, and result $1$ (and state
$\ket{-})$) with probability
$\frac{1}{2}$, because of the representation of $\ket{0}$ in the
diagonal basis noted above. If a classical bit is represented by a
qubit using either the standard or diagonal basis, then a measurement
with respect to the correct basis results in the original bit, but a
measurement with respect to the other basis results in $0$ or $1$ with equal
probability. This behaviour is used by the quantum bit-commitment
protocol which we discuss in Section~\ref{sec-bitcommitment}.  

To go beyond single-qubit systems, we consider tensor products of
spaces (in contrast to the cartesian products used in classical
systems). If spaces $U$ and $V$ have bases $\{u_i\}$ and $\{v_j\}$
then $U\otimes V$ has basis $\{u_i\otimes v_j\}$. In particular, a
system consisting of $n$ qubits has a $2^n$-dimensional space whose
standard basis is $\ket{00\ldots 0}\ldots\ket{11\ldots 1}$. We can now
consider measurements of single qubits or collective measurements of
multiple qubits. For example, a $2$-qubit system has basis
$\ket{00},\ket{01},\ket{10},\ket{11}$ and a general state is
$\alpha\ket{00}+\beta\ket{01}+\gamma\ket{10}+\delta\ket{11}$ with
$\ms{\alpha}+\ms{\beta}+\ms{\gamma}+\ms{\delta}=1$. Measuring the
first qubit gives result $0$ with probability $\ms{\alpha}+\ms{\beta}$
(leaving the system in state
$\frac{1}{\ms{\alpha}+\ms{\beta}}(\alpha\ket{00}+\beta\ket{01})$) and
result $1$ with probability $\ms{\gamma}+\ms{\delta}$ (leaving the
system in state
$\frac{1}{\ms{\gamma}+\ms{\delta}}(\gamma\ket{10}+\delta\ket{11})$).
Measuring both qubits simultaneously gives result $0$ with probability
$\ms{\alpha}$ (leaving the system in state $\ket{00}$), result $1$
with probability $\ms{\beta}$ (leaving the system in state $\ket{01}$)
and so on; note that the association of basis states
$\ket{00},\ket{01},\ket{10},\ket{11}$ with results $0,1,2,3$ is just a
conventional choice. The power of quantum computing, in an algorithmic
sense, results from calculating with superpositions of states; all the
states are transformed simultaneously (\emph{quantum parallelism}) and
the effect increases exponentially with the dimension of the state
space. The challenge in quantum algorithm design is to make
measurements which enable this parallelism to be exploited; in general
this is very difficult.

We will make use of the \emph{conditional not} ($\qgate{CNot}$)
transformation on pairs of qubits. Its action on basis states is
defined by
\[
\begin{array}{@{\extracolsep{1mm}}cccc}
\ket{00}\mapsto\ket{00} & \ket{01}\mapsto\ket{01} &
\ket{10}\mapsto\ket{11} & \ket{11}\mapsto\ket{10} 
\end{array}
\]
which can be understood as inverting the second qubit if and only if
the first qubit is set, although in general we need to consider the
effect on non-basis states.

Systems of two or more qubits can exhibit the phenomenon of
\emph{entanglement}, meaning that the states of the qubits are
correlated. For example, consider a measurement of the first qubit of
the state $\frac{1}{\sqrt{2}}(\ket{00}+\ket{11})$. The result is $0$
(and resulting state $\ket{00}$) with probability $\frac{1}{2}$, or
$1$ (and resulting state $\ket{11}$) with probability
$\frac{1}{2}$. In either case a subsequent measurement of the second
qubit gives a definite (non-probabilistic) result which is always the
same as the result of the first measurement. This is true even if the
entangled qubits are physically separated.  Entanglement illustrates
the key difference between the use of tensor product (in quantum
systems) and cartesian product (in classical systems): an entangled
state of two qubits is one which cannot be decomposed as a pair of
single-qubit states. Entanglement is used in an essential way in the
quantum teleportation protocol which we discuss in
Section~\ref{sec-teleportation}. That example uses the
$\qgate{CNot}$ transformation to create entanglement:
$\qgate{CNot}((\qgate{H}\otimes I)\ket{00}) =
\frac{1}{\sqrt{2}}(\ket{00}+\ket{11})$.

\section{Examples of Modelling in CQP}
\label{sec-examples}
\subsection{A Quantum Coin-Flipping Game}
\label{sec-coinflip}
Our first example is based on a scenario used by Meyer
\cite{MeyerDA:quas} to initiate the study of quantum game
theory. Players $P$ and $Q$ play the following game: $P$ places a
coin, head upwards, in a box, and then the players take turns ($Q$,
then $P$, then $Q$) to optionally turn the coin over, without being
able to see it. Finally the box is opened and $Q$ wins if the coin is
head upwards.

Clearly neither player has a winning strategy, but the situation
changes if the coin is a quantum system, represented by a qubit
($\ket{0}$ for head upwards, $\ket{1}$ for tail upwards). Turning the
coin over corresponds to the transformation $\sigma_1$, and this is
what $P$ can do. But suppose that $Q$ can apply $\qgate{H}$, which
corresponds to transforming from head upwards ($\ket{0}$) to a
superposition of head upwards and tail upwards
($\frac{1}{\sqrt{2}}(\ket{0}+\ket{1})$), and does this on both
turns. Then we have two possible runs of the game, (a) and (b):
\begin{center}
\renewcommand{\arraystretch}{1.2}
\begin{tabular}{l|l}
\multicolumn{2}{c}{(a)}\\
Action & State \\ \hline
& $\ket{0}$ \\
$Q$: $\qgate{H}$ & $\frac{1}{\sqrt{2}}(\ket{0}+\ket{1})$ \\
$P$: $\sigma_1$ & $\frac{1}{\sqrt{2}}(\ket{1}+\ket{0})$ \\
$Q$: $\qgate{H}$ & $\ket{0}$
\end{tabular}\hspace{5mm}
\begin{tabular}{l|l}
\multicolumn{2}{c}{(b)}\\
Action & State \\ \hline
\phantom{$P$: $\sigma_1$} & $\ket{0}$ \\
$Q$: $\qgate{H}$ & $\frac{1}{\sqrt{2}}(\ket{0}+\ket{1})$ \\
$P$: $-$ & $\frac{1}{\sqrt{2}}(\ket{0}+\ket{1})$ \\
$Q$: $\qgate{H}$ & $\ket{0}$
\end{tabular}
\renewcommand{\arraystretch}{1}
\end{center}
and in each case the coin finishes head upwards. To verify this we
calculate that the state $\frac{1}{\sqrt{2}}(\ket{0}+\ket{1})$ is
invariant under $\sigma_1$:
\[
\matr{0}{1}{1}{0}\frac{1}{\sqrt{2}}\vect{1}{1} = \frac{1}{\sqrt{2}}\vect{1}{1}
\]
and that the Hadamard transformation $\qgate{H}$ is self-inverse:
\[
\frac{1}{\sqrt{2}}\matr{1}{1}{1}{-1}\frac{1}{\sqrt{2}}\matr{1}{1}{1}{-1} = \matr{1}{0}{0}{1}
\]

Meyer considers game-theoretic issues relating to the expected outcome
of repeated runs, but we just model a single run in CQP
(Figure~\ref{fig-coinflip}). Most of the syntax of CQP is based on
typed pi-calculus, using fairly common notation (for example, see
Pierce and Sangiorgi's presentation \cite{PierceBC:typsmpfull}). $P$
and $Q$ communicate by means of the typed channel
$\tid{s}{\Chant{\Qbit}}$ which carries qubits. It is a parameter of
both $P$ and $Q$. At the top level, $\pname{System}$ creates $s$ with
$(\new
\tid{s}{\Chant{\Qbit}})$ and starts $P$ and $Q$ in parallel. $Q$ and
$\pname{System}$
are also parameterized by $x$, the qubit representing the initial
state of the coin.

$Q$ applies ($\trans{x}{\qgate{H}}$) the Hadamard transformation to
$x$; this syntax is based on Selinger's QPL
\cite{SelingerP:towqpl}. This expression is converted into an action
by $\action{\ldots}$. Using a standard pi-calculus programming style,
$Q$ creates a channel $t$ and sends ($\outp{s}{x,t}$) it to $P$ along
with the qubit $x$. $P$ will use $t$ to send the qubit back, and $Q$
receives it with $\inp{t}{\tid{z}{\Qbit}}$, binding it to the name $z$
in the rest of the code. Finally $Q$ applies $\qgate{H}$ again, and
continues with some behaviour $C(z)$.

$P$ contains two branches of behaviour, corresponding to the
possibilities of applying (second branch) or not applying (first
branch) the transformation $\sigma_1$. Both branches terminate with
the null process $\nil$. The branches are placed in
parallel\footnote{Simpler definitions can be obtained if we add
guarded sums to CQP; there is then no need for the channel $t$. This
is straightforward but we have chosen instead to simplify the
presentation of the semantics.}  and the operational semantics means
that only one of them interacts with $Q$; the other is effectively
$\pname{Garbage}$ (different in each case).

Figure~\ref{fig-coinflip-execution} shows the execution (combining some
steps) of
$\pname{System}$ according to the operational semantics which we will
define formally in Section~\ref{sec-syntax}. Reduction takes place on
configurations $\cnfig{\sigma}{\phi}{P}$ where $\sigma$ is a list of
qubits and their collective state, $\phi$ lists the channels which
have been created, and $P$ is a process term. Note that the state of
the qubits \emph{must} be a global property in order to be physically
realistic. We record the channels globally in order to give the
semantics a uniform style; this is different from the usual approach
to pi-calculus semantics, but (modulo garbage collection) is
equivalent to expanding the scope of every \textsf{new} before
beginning execution.

The execution of $\pname{System}$ tracks the informal calculation
which we worked through above. Our CQP model makes the
manipulation of the qubit very explicit; there are other ways to
express the behaviour (including putting everything into a single
process with no communication), but the point is that we have a
framework in which to discuss such issues.

\begin{figure*}
\[
\begin{array}{lcl}
\pname{P}(\tid{s}{\Chant{\Qbit}}) & = &
\inp{s}{\tid{y}{\Qbit},\tid{t}{\Chant{\Qbit}}}\sep\outp{t}{y}\sep\nil\\
& \parallel &
\inp{s}{\tid{y}{\Qbit},\tid{t}{\Chant{\Qbit}}}\sep\action{\trans{y}{\sigma_1}}\sep\outp{t}{y}\sep\nil
\\ \\
\multicolumn{3}{l}{\pname{Q}(\tid{x}{\Qbit},\tid{s}{\Chant{\Qbit}}) =
\action{\trans{x}{\qgate{H}}}\sep(\new\tid{t}{\Chant{\Qbit}})(\outp{s}{x,t}\sep\inp{t}{\tid{z}{\Qbit}}\sep\action{\trans{z}{\qgate{H}}}\sep\pname{C}(z))}
\\ \\
\multicolumn{3}{l}{\pname{System}(\tid{x}{\Qbit}) = (\new\tid{s}{\Chant{\Qbit}})(\pname{P}(s)\parallel\pname{Q}(x,s))}
\end{array}
\]
\caption{The quantum coin-flipping game in CQP}
\label{fig-coinflip}
\end{figure*}

\begin{figure*}
\[
\begin{array}{ccc}
\multicolumn{3}{c}{x = \ket{0}\,;\emptyset\,;\pname{System}(x)}
\\
& \downarrow & \text{\scriptsize expand definition} \\
\multicolumn{3}{c}{x = \ket{0}\,;\emptyset\,;
(\new\tid{s}{\Chant{\Qbit}})(\pname{P}(s)\parallel\pname{Q}(x,s))}
\\
& \downarrow & \text{\scriptsize create channel $s$}\\
\multicolumn{3}{c}{x = \ket{0}\,;s\,;
\pname{P}(s)\parallel\pname{Q}(x,s)}
\\
& \downarrow & \text{\scriptsize expand definitions} \\
\multicolumn{3}{c}{x = \ket{0}\,;s\,;} \\
\multicolumn{3}{c}{\inp{s}{\tid{y}{\Qbit},\tid{t}{\Chant{\Qbit}}}\sep\outp{t}{y}\sep\nil
\,\parallel\,
\inp{s}{\tid{y}{\Qbit},\tid{t}{\Chant{\Qbit}}}\sep\action{\trans{y}{\sigma_1}}\sep\outp{t}{y}\sep\nil}\\
\multicolumn{3}{c}{\parallel\,
\action{\trans{x}{\qgate{H}}}\sep(\new\tid{t}{\Chant{\Qbit}})(\outp{s}{x,t}\sep\inp{t}{\tid{z}{\Qbit}}\sep\action{\trans{z}{\qgate{H}}}\sep\pname{C}(z)}
\\
& \downarrow & \text{\scriptsize transform $x$} \\
\multicolumn{3}{c}{x = \frac{1}{\sqrt{2}}(\ket{0}+\ket{1})\,;s\,;} \\
\multicolumn{3}{c}{\inp{s}{\tid{y}{\Qbit},\tid{t}{\Chant{\Qbit}}}\sep\outp{t}{y}\sep\nil
\,\parallel\,
\inp{s}{\tid{y}{\Qbit},\tid{t}{\Chant{\Qbit}}}\sep\action{\trans{y}{\sigma_1}}\sep\outp{t}{y}\sep\nil}\\
\multicolumn{3}{c}{\parallel\,
(\new\tid{t}{\Chant{\Qbit}})(\outp{s}{x,t}\sep\inp{t}{\tid{z}{\Qbit}}\sep\action{\trans{z}{\qgate{H}}}\sep\pname{C}(z))}
\\
& \downarrow & \text{\scriptsize create channel $t$}\\
\multicolumn{3}{c}{x = \frac{1}{\sqrt{2}}(\ket{0}+\ket{1})\,;s,t\,;} \\
\multicolumn{3}{c}{\inp{s}{\tid{y}{\Qbit},\tid{t}{\Chant{\Qbit}}}\sep\outp{t}{y}\sep\nil
\,\parallel\,
\inp{s}{\tid{y}{\Qbit},\tid{t}{\Chant{\Qbit}}}\sep\action{\trans{y}{\sigma_1}}\sep\outp{t}{y}\sep\nil}\\
\multicolumn{3}{c}{\parallel\,
\outp{s}{x,t}\sep\inp{t}{\tid{z}{\Qbit}}\sep\action{\trans{z}{\qgate{H}}}\sep\pname{C}(z)}\\
\swarrow & & \phantom{\text{\scriptsize communication}} \searrow \text{\scriptsize communication}\\
x = \frac{1}{\sqrt{2}}(\ket{0}+\ket{1})\,;s,t\,; & &
x = \frac{1}{\sqrt{2}}(\ket{0}+\ket{1})\,;s,t\,; \\
\outp{t}{x}\sep\nil
\,\parallel\,
\pname{Garbage}
& & 
\pname{Garbage}
\,\parallel\,
\action{\trans{x}{\sigma_1}}\sep\outp{t}{x}\sep\nil
\\
\parallel\,
\inp{t}{\tid{z}{\Qbit}}\sep\action{\trans{z}{\qgate{H}}}\sep\pname{C}(z)
& & 
\parallel\,
\inp{t}{\tid{z}{\Qbit}}\sep\action{\trans{z}{\qgate{H}}}\sep\pname{C}(z)\\
\downarrow & & \phantom{\text{\scriptsize transform $x$}} \downarrow \text{\scriptsize transform $x$}\\
x = \frac{1}{\sqrt{2}}(\ket{0}+\ket{1})\,;s,t\,; & &
x = \frac{1}{\sqrt{2}}(\ket{0}+\ket{1})\,;s,t\,; \\
\nil
\,\parallel\,
\pname{Garbage}\parallel\,
\action{\trans{x}{\qgate{H}}}\sep\pname{C}(x)
& & 
\pname{Garbage}
\,\parallel\,
\outp{t}{x}\sep\nil
\\
& & 
\parallel\,
\inp{t}{\tid{z}{\Qbit}}\sep\action{\trans{z}{\qgate{H}}}\sep\pname{C}(z)
\\
\downarrow & & \phantom{\text{\scriptsize communication}} \downarrow \text{\scriptsize communication}\\
x = \ket{0}\,;s,t\,;\pname{Garbage}\parallel\,\pname{C}(x) & &
x = \frac{1}{\sqrt{2}}(\ket{0}+\ket{1})\,;s,t\,; \\
& &
\pname{Garbage}\parallel\,\nil\parallel\,\action{\trans{x}{\qgate{H}}}\sep\pname{C}(x) 
\\
& & \phantom{\text{\scriptsize transform $x$}} \downarrow \text{\scriptsize transform $x$}\\
& & x = \ket{0}\,;s,t\,;\pname{Garbage}\parallel\,\pname{C}(x)
\end{array}
\]
\caption{Execution of the coin-flipping game}
\label{fig-coinflip-execution}
\end{figure*}

\begin{figure*}
\[
\begin{array}{l}
\pname{Alice}(\tid{x}{\Qbit},\tid{c}{\Chant{\ranget{0}{3}}},\tid{z}{\Qbit})
= 
\action{\trans{z,x}{\qgate{CNot}}}\sep\action{\trans{z}{\qgate{H}}}\sep\outp{c}{\measure
z,x}\sep\nil \\ \\
\pname{Bob}(\tid{y}{\Qbit},\tid{c}{\Chant{\ranget{0}{3}}})
= \inp{c}{\tid{r}{\ranget{0}{3}}}\sep\action{\trans{y}{\sigma_{r}}}\sep\pname{Use}(y)
\\ \\
\pname{System}(\tid{x}{\Qbit},\tid{y}{\Qbit},\tid{z}{\Qbit})
= (\new \tid{c}{\chant{\ranget{0}{3}}})(\pname{Alice}(x,c,z) \parallel \pname{Bob}(y,c))
\end{array}
\]
\caption{Quantum teleportation in CQP}
\label{fig-teleportation}
\end{figure*}

\begin{figure*}
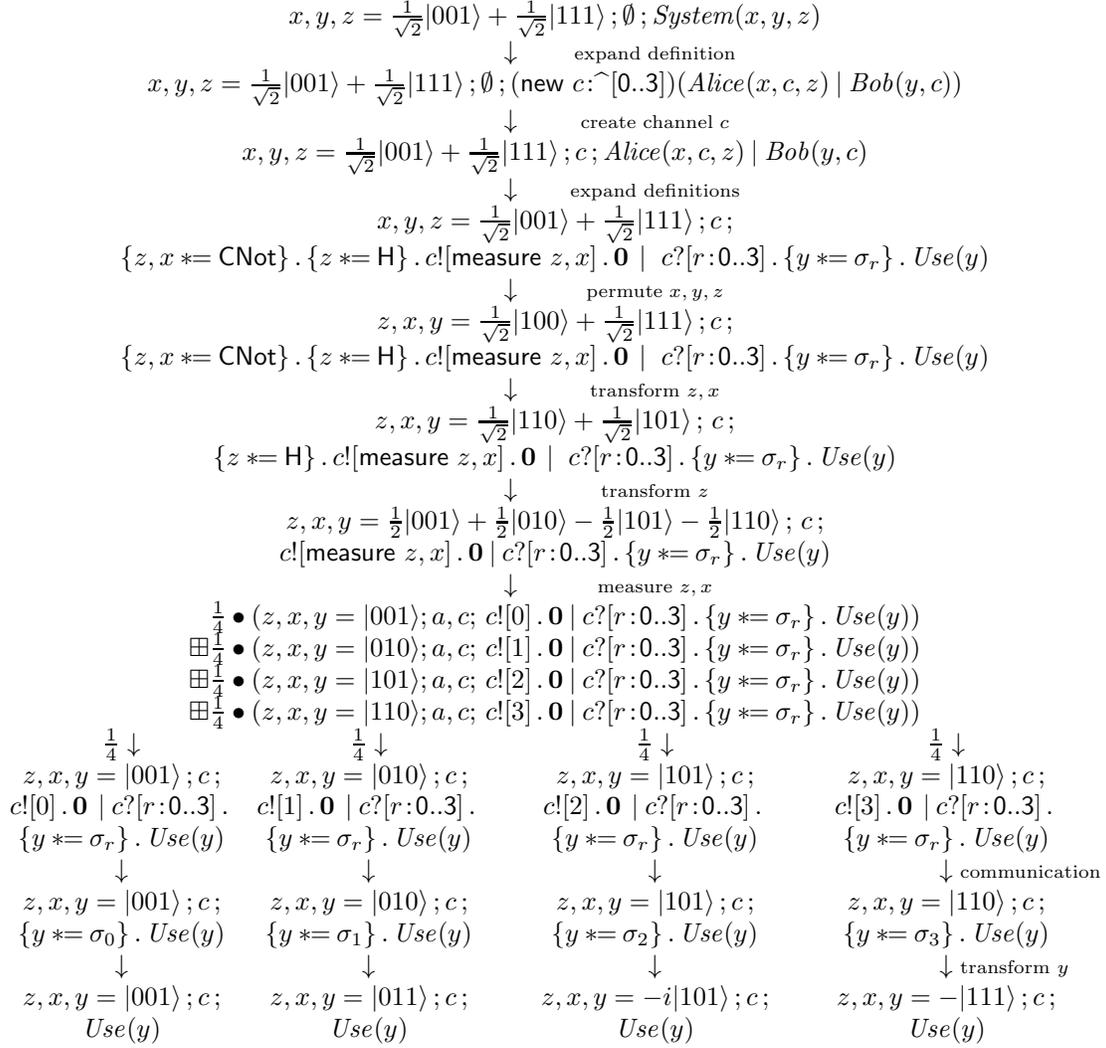

\[
\begin{array}{ccccc}
\multicolumn{5}{c}{x,y,z = \frac{1}{\sqrt{2}}\ket{001}+\frac{1}{\sqrt{2}}\ket{111}\,;\emptyset\,;\pname{System}(x,y,z)}
\\
& & \downarrow & \text{\scriptsize{expand definition}} \\
\multicolumn{5}{c}{x,y,z = \frac{1}{\sqrt{2}}\ket{001}+\frac{1}{\sqrt{2}}\ket{111}\,;\emptyset\,;
(\new
\tid{c}{\Chant{\ranget{0}{3}}})(\pname{Alice}(x,c,z)\parallel\pname{Bob}(y,c))}
\\
& & \downarrow & \text{\scriptsize{create channel $c$}} \\
\multicolumn{5}{c}{x,y,z = \frac{1}{\sqrt{2}}\ket{001}+\frac{1}{\sqrt{2}}\ket{111}\,;c\,; 
\pname{Alice}(x,c,z)\parallel\pname{Bob}(y,c)}
\\
& & \downarrow & \text{\scriptsize{expand definitions}} \\
\multicolumn{5}{c}{x,y,z = \frac{1}{\sqrt{2}}\ket{001}+\frac{1}{\sqrt{2}}\ket{111}\,;c\,;} \\
\multicolumn{5}{c}{\action{\trans{z,x}{\qgate{CNot}}}\sep\action{\trans{z}{\qgate{H}}}\sep\outp{c}{\measure
z,x}\sep\nil\,
\parallel\ 
\inp{c}{\tid{r}{\ranget{0}{3}}}\sep\action{\trans{y}{\sigma_{r}}}\sep\pname{Use}(y)}
\\
& & \downarrow & \text{\scriptsize{permute $x,y,z$}} \\
\multicolumn{5}{c}{z,x,y = \frac{1}{\sqrt{2}}\ket{100}+\frac{1}{\sqrt{2}}\ket{111}\,;c\,;} \\
\multicolumn{5}{c}{\action{\trans{z,x}{\qgate{CNot}}}\sep\action{\trans{z}{\qgate{H}}}\sep\outp{c}{\measure
z,x}\sep\nil\,
\parallel\ 
\inp{c}{\tid{r}{\ranget{0}{3}}}\sep\action{\trans{y}{\sigma_{r}}}\sep\pname{Use}(y)}
\\
& & \downarrow & \text{\scriptsize{transform $z,x$}} \\
\multicolumn{5}{c}{z,x,y = \frac{1}{\sqrt{2}}\ket{110}+\frac{1}{\sqrt{2}}\ket{101}\,;\,c\,;} \\
\multicolumn{5}{c}{\action{\trans{z}{\qgate{H}}}\sep\outp{c}{\measure
z,x}\sep\nil\,
\parallel\ 
\inp{c}{\tid{r}{\ranget{0}{3}}}\sep\action{\trans{y}{\sigma_{r}}}\sep\pname{Use}(y)}
\\
& & \downarrow & \text{\scriptsize{transform $z$}} \\
\multicolumn{5}{c}{z,x,y = \frac{1}{2}\ket{001}+\frac{1}{2}\ket{010}-\frac{1}{2}\ket{101}-\frac{1}{2}\ket{110}\,;\,c\,;} \\
\multicolumn{5}{c}{\outp{c}{\measure z,x}\sep\nil
\parallel
\inp{c}{\tid{r}{\ranget{0}{3}}}\sep\action{\trans{y}{\sigma_{r}}}\sep\pname{Use}(y)}
\\
& & \downarrow & \text{\scriptsize{measure $z,x$}} \\
\multicolumn{5}{c}{\phantom{\bp}\prob{\frac{1}{4}}{\cnfig{z,x,y =
\ket{001}}{a,c}{\,\outp{c}{0}\sep\nil\parallel\inp{c}{\tid{r}{\ranget{0}{3}}}\sep\action{\trans{y}{\sigma_{r}}}\sep\pname{Use}(y)}}}\\
\multicolumn{5}{c}{\bp\prob{\frac{1}{4}}{\cnfig{z,x,y =
\ket{010}}{a,c}{\,\outp{c}{1}\sep\nil\parallel\inp{c}{\tid{r}{\ranget{0}{3}}}\sep\action{\trans{y}{\sigma_{r}}}\sep\pname{Use}(y)}}}\\
\multicolumn{5}{c}{\bp\prob{\frac{1}{4}}{\cnfig{z,x,y =
\ket{101}}{a,c}{\,\outp{c}{2}\sep\nil\parallel\inp{c}{\tid{r}{\ranget{0}{3}}}\sep\action{\trans{y}{\sigma_{r}}}\sep\pname{Use}(y)}}}\\
\multicolumn{5}{c}{\bp\prob{\frac{1}{4}}{\cnfig{z,x,y =
\ket{110}}{a,c}{\,\outp{c}{3}\sep\nil\parallel\inp{c}{\tid{r}{\ranget{0}{3}}}\sep\action{\trans{y}{\sigma_{r}}}\sep\pname{Use}(y)}}}
\\
\frac{1}{4}\downarrow & \frac{1}{4}\downarrow & & \frac{1}{4}\downarrow
& \frac{1}{4}\downarrow \\
z,x,y = \ket{001}\,;c\,; &
z,x,y = \ket{010}\,;c\,; & &
z,x,y = \ket{101}\,;c\,; &
z,x,y = \ket{110}\,;c\,;
\\
\outp{c}{0}\sep\nil\,\parallel\inp{c}{\tid{r}{\ranget{0}{3}}}\sep
&
\outp{c}{1}\sep\nil\,\parallel\inp{c}{\tid{r}{\ranget{0}{3}}}\sep
& &
\outp{c}{2}\sep\nil\,\parallel\inp{c}{\tid{r}{\ranget{0}{3}}}\sep
&
\outp{c}{3}\sep\nil\,\parallel\inp{c}{\tid{r}{\ranget{0}{3}}}\sep
\\
\action{\trans{y}{\sigma_{r}}}\sep\pname{Use}(y)
&
\action{\trans{y}{\sigma_{r}}}\sep\pname{Use}(y)
& &
\action{\trans{y}{\sigma_{r}}}\sep\pname{Use}(y)
&
\action{\trans{y}{\sigma_{r}}}\sep\pname{Use}(y)
\\
\downarrow & \downarrow & & \downarrow & \phantom{\text{\scriptsize communication}} \downarrow \text{\scriptsize communication}
\\
z,x,y = \ket{001}\,;c\,; &
z,x,y = \ket{010}\,;c\,; & &
z,x,y = \ket{101}\,;c\,; &
z,x,y = \ket{110}\,;c\,;
\\
\action{\trans{y}{\sigma_{0}}}\sep\pname{Use}(y)
&
\action{\trans{y}{\sigma_{1}}}\sep\pname{Use}(y)
& &
\action{\trans{y}{\sigma_{2}}}\sep\pname{Use}(y)
&
\action{\trans{y}{\sigma_{3}}}\sep\pname{Use}(y)
\\
\downarrow & \downarrow & & \downarrow & \phantom{\text{\scriptsize
transform $y$}} \downarrow \text{\scriptsize
transform $y$}
\\
z,x,y = \ket{001}\,;c\,; &
z,x,y = \ket{011}\,;c\,; & &
z,x,y = -i\ket{101}\,;c\,; &
z,x,y = -\ket{111}\,;c\,;
\\
\pname{Use}(y) & \pname{Use}(y) & & \pname{Use}(y) & \pname{Use}(y)
\end{array}
\]
\caption{Execution of the quantum teleportation protocol}
\label{fig-teleportation-execution}
\end{figure*}

\begin{figure*}
\[
\begin{array}{lll}
\multicolumn{3}{l}{\pname{Alice}'(\tid{s}{\Chant{\Qbit}},\tid{c}{\Chant{\ranget{0}{3}}},\tid{z}{\Qbit})
= \inp{s}{\tid{x}{\Qbit}}\sep\pname{Alice}(x,c,a)} \\ \\
\multicolumn{3}{l}{\pname{Bob}'(\tid{t}{\Chant{\Qbit}},\tid{c}{\Chant{\ranget{0}{3}}}) = 
\inp{t}{\tid{y}{\Qbit}}\sep\pname{Bob}(y,c)} \\ \\
\multicolumn{3}{l}{\pname{Source}(\tid{s}{\Chant{\Qbit}},\tid{t}{\Chant{\Qbit}})
= (\qbit x,y)(\action{\trans{x}{\qgate{H}}}\sep\action{\trans{x,y}{\qgate{CNot}}}\sep
\outp{s}{x}\sep\outp{t}{y}\sep\nil)} \\ \\
\multicolumn{3}{l}{\pname{System}'(\tid{z}{\Qbit}) = (\new \tid{c}{\Chant{\ranget{0}{3}}}, \tid{s}{\Chant{\Qbit}},
\tid{t}{\Chant{\Qbit}})(\pname{Alice}'(s,c,z) \parallel
\pname{Bob}'(t,c) \parallel \pname{Source}(s,t))}
\end{array}
\]
\caption{Quantum teleportation with an EPR source}
\label{fig-teleportation-source}
\end{figure*}

\begin{figure*}
\[
\begin{array}{lll}
\multicolumn{3}{l}{\pname{Alice}(\tid{x}{\Bit},\tid{xs}{\Listt{\Bit}},\tid{c}{\Chant{\Qbit}},\tid{d}{\Chant{\Bit}},\tid{e}{\Chant{\Int}},\tid{f}{\Chant{\Listt{\Bit}}})
=}\\
\hspace{5mm} & \multicolumn{2}{l}{\outp{e}{\kw{length}(xs)}\sep\pname{AliceSend}(x,\kw{length}(xs),xs,xs,c,d,e,f)}
\\ \\
\multicolumn{3}{l}{\pname{AliceSend}(\tid{x}{\Bit},\tid{n}{\Int},\tid{xs}{\Listt{\Bit}},\tid{ys}{\Listt{\Bit}},\tid{c}{\Chant{\Qbit}},\tid{d}{\Chant{\Bit}},\tid{e}{\Chant{\Int}},\tid{f}{\Chant{\Listt{\Bit}}})
=}\\
& \multicolumn{2}{l}{\kw{if}\ n=0\ \kw{then}\ \pname{AliceReceive}(x,\kw{length}(ys),ys,c,d,e,f)}\\
& \kw{else}\ (\qbit q)( \hspace{-2mm}& \action{\kw{if}\ \kw{hd}(xs) = 1\ \kw{then}\
\trans{q}{\sigma_x}\ \kw{else}\ \unit} \sep \action{\kw{if}\ x=1\
\kw{then}\ \trans{q}{\qgate{H}}\ \kw{else}\ \unit} \sep \outp{c}{q} \sep\\
& & \pname{AliceSend}(x,n-1,\kw{tl}(xs),ys,c,d,e,f))
\\ \\
\multicolumn{3}{l}{\pname{AliceReceive}(\tid{x}{\Bit},\tid{n}{\Int},\tid{ys}{\Listt{\Bit}},\tid{d}{\Chant{\Bit}},\tid{f}{\Chant{\Listt{\Bit}}})
= \inp{d}{\tid{g}{\Bit}}\sep\outp{d}{x}\sep\outp{f}{ys}\sep\nil}
\\ \\
\multicolumn{3}{l}{\pname{Bob}(\tid{c}{\Chant{\Qbit}},\tid{d}{\Chant{\Bit}},\tid{e}{\Chant{\Int}},\tid{f}{\Chant{\Listt{\Bit}}},\tid{r}{\Chant{\Bit}})
= \inp{e}{\tid{n}{\Int}}\sep\pname{BobReceive}([\,],n,c,d,f,r)}
\\ \\
\multicolumn{3}{l}{\pname{BobReceive}(\tid{m}{\Listt{(\Prodt{\Bit}{\Bit})}},\tid{n}{\Int},\tid{c}{\Chant{\Qbit}},\tid{d}{\Chant{\Bit}},\tid{e}{\Chant{\Int}},\tid{f}{\Chant{\Listt{\Bit}}},\tid{r}{\Chant{\Bit}})=}\\
& \multicolumn{2}{l}{\kw{if}\ n=0\ \kw{then}\
\inp{r}{\tid{g}{\Bit}}\sep\outp{d}{g}\sep\inp{d}{\tid{a}{\Bit}}\sep\inp{f}{\tid{vs}{\Listt{\Bit}}}\sep\pname{BobVerify}(m,vs,a,\kw{length}(m))}\\
& \multicolumn{2}{l}{\kw{else}\
\inp{c}{\tid{x}{\Qbit}}\sep\inp{r}{\tid{y}{\Bit}}\sep\action{\kw{if}\
y=1\ \kw{then}\ \trans{x}{\qgate{H}}\ \kw{else}\
\unit}\sep\pname{BobReceive}(m@[(y,\msure{x})],n-1,c,d,r)}
\\ \\
\multicolumn{3}{l}{\pname{BobVerify}(\tid{m}{\Listt{(\Prodt{\Bit}{\Bit})}},\tid{vs}{\Listt{\Bit}},\tid{a}{\Bit},\tid{n}{\Int})
=}\\
& \multicolumn{2}{l}{\kw{if}\ n=0\ \kw{then}\ \pname{Verified}}\\
& \multicolumn{2}{l}{\kw{else}\ \kw{if}\ \kw{fst}(\kw{hd}(m))=a\
\kw{then}} \\
& & \kw{if}\ \kw{snd}(\kw{hd}(m))=\kw{hd}(vs)\ \kw{then}\
\pname{BobVerify}(\kw{tl}(m),\kw{tl}(vs),a,n-1)\\
& & \kw{else}\ \pname{NotVerified} \\
& \multicolumn{2}{l}{\kw{else}\
\pname{BobVerify}(\kw{tl}(m),\kw{tl}(vs),a,n-1)}
\\ \\
\multicolumn{3}{l}{\pname{Random}(\tid{r}{\Chant{\Bit}}) = (\qbit
q)(\action{\trans{q}{\qgate{H}}}\sep\outp{r}{\msure{q}}\sep\pname{Random(r)})}
\\ \\
\multicolumn{3}{l}{\pname{System}(\tid{x}{\Bit},\tid{xs}{\Listt{\Bit}})
=}\\
& \multicolumn{2}{l}{(\new
\tid{c}{\Chant{\Qbit}},\tid{d}{\Chant{\Bit}},\tid{e}{\Chant{\Int}},\tid{f}{\Chant{\Listt{\Bit}}},\tid{r}{\Chant{\Bit}})}\\
& & (\pname{Alice}(x,xs,c,d,e,f) \parallel \pname{Bob}(c,d,e,f,r)
\parallel \pname{Random}(r))
\end{array}
\]
\caption{Quantum bit-commitment in CQP}
\label{fig-bitcommitment}
\end{figure*}

\begin{figure*}
\[
\begin{array}{rcl}
T & \bnf & \Int \alt \Unit \alt \Qbit \alt \Chant{\vec{T}} \alt \Op{1} \alt \Op{2} \alt
\ldots \\
v & \bnf & x \alt 
\num{0} \alt \num{1} \alt \ldots \alt \unit \alt \qgate{H} \alt \ldots \\
e & \bnf & v \alt \msure{\vec{e}} \alt \trans{\vec{e}}{e} \alt \plus{e}{e} \\
P & \bnf & \nil \alt (P \parallel P) \alt
\inp{e}{\tid{\vec{x}}{\vec{T}}}\sep P \alt \outp{e}{\vec{e}}\sep P
\alt \action{e}\sep P \alt (\new\tid{x}{T})P \alt (\qbit x)P
\end{array}
\]
\caption{Syntax of CQP}
\label{fig-syntax}
\end{figure*}

\begin{figure*}
\[
\begin{array}{rcl}
v & \bnf & \ldots \alt q \alt c \\
E & \bnf & \ctxt{}{~} \alt \msure{E,\vec{e}} \alt
\msure{v,E,\vec{e}} \alt \ldots \alt \msure{\vec{v},E} \alt \trans{E,\vec{e}}{e} \alt
\trans{v,E,\vec{e}}{e} \\
& & \alt \ldots \alt \trans{\vec{v}}{E} \alt
\plus{E}{e} \alt \plus{v}{E} \\
F & \bnf & 
\inp{\ctxt{}{~}}{\tid{\vec{x}}{\vec{T}}}\sep P \alt \outp{\ctxt{}{~}}{\vec{e}}\sep P
\alt \outp{v}{\ctxt{}{~},\vec{e}}\sep P \alt \outp{v}{v,\ctxt{}{~},\vec{e}}\sep P \alt
\ldots \alt \outp{v}{\vec{v},\ctxt{}{~}}\sep P \alt \action{\ctxt{}{~}}\sep P
\end{array}
\]
\caption{Internal syntax of CQP}
\label{fig-internal-syntax}
\end{figure*}

\begin{figure*}
\[
\begin{array}{c@{\extracolsep{10mm}}cc}
P \parallel \nil \scong P
&
P \parallel Q \scong Q \parallel P
&
P\parallel(Q\parallel R) \scong (P\parallel Q)\parallel R
\\ \\
(\Snil)
&
(\Scomm)
&
(\Sassoc)
\end{array}
\]
\caption{Structural congruence}
\label{fig-structural-congruence}
\end{figure*}

\subsection{Quantum Teleportation}
\label{sec-teleportation}
The quantum teleportation protocol \cite{BennettCH:teluqs} is a
procedure for transmitting a quantum state via a non-quantum
medium. This protocol is particularly important: not only is it a
fundamental component of several more complex protocols, but it is
likely to be a key enabling technology for the development of the
\emph{quantum repeaters}
\cite{deRiedmattenH:londqt} which will be necessary in
large-scale quantum communication networks.

Figure~\ref{fig-teleportation} shows a simple model of the quantum
teleportation protocol. Alice and Bob each possess one qubit ($x$ for
Alice, $y$ for Bob) of an
entangled pair whose state is
$\frac{1}{\sqrt{2}}(\ket{00} + \ket{11})$. At this point
we are assuming that appropriate qubits will be supplied to Alice and
Bob as parameters of the system. Alice is also parameterized by a
qubit $z$, whose state is to be teleported. She applies
($\trans{z,x}{\qgate{CNot}}$) the conditional not
transformation to $z$ and $x$ and then applies
($\trans{z}{\qgate{H}}$) the
Hadamard transformation to $z$, finally measuring
$z$ and $x$ to yield a two-bit classical value which she sends
($\outp{c}{\measure z,x}$) to Bob on the typed
channel $\tid{c}{\Chant{\ranget{0}{3}}}$ and then terminates
($\nil$). Bob receives ($\inp{c}{\tid{r}{\ranget{0}{3}}}$) this value
and uses it to select\footnote{We can easily extend the expression
language of CQP to allow explicit testing of $r$.} a \emph{Pauli} transformation
$\sigma_{0}\ldots\sigma_{3}$ to apply ($\trans{y}{\sigma_{r}}$) to
$y$. The result is that Bob's qubit $y$ takes on the state of $z$,
without a physical qubit having been transmitted from Alice to
Bob. Bob may then use $y$ in his continuation process
$\pname{Use}(y)$. 

This example introduces measurement, with a syntax similar to that of
Selinger's QPL \cite{SelingerP:towqpl}. We treat measurement as an
expression, executed for its value as well as its side-effect on the
quantum state. Because the result of a measurement is probabilistic,
evaluation of a $\kw{measure}$ expression introduces a probability
distribution over configurations: $\bp_{0\leqslant i\leqslant
n}\,\prob{p_i}{\cnfig{\sigma_i}{\phi_i}{P_i}}$. The next step is a
probabilistic transition to one of the configurations; no reduction
takes place underneath a probability distribution. In general a
configuration reduces non-deterministically to
one of a collection of probability distributions over configurations
(in some cases this is trivial, with only one distribution or only one
configuration within a distribution). A non-trivial probability
distribution makes a probabilistic transition to a
single configuration; this step is omitted in the case of a trivial
distribution.

Figure~\ref{fig-teleportation-execution} shows the complete execution
of $\pname{System}$ in the particular case in which $z$, the qubit
being teleported, has state $\ket{1}$. The measurement produces a
probability distribution over four configurations, but in all cases
the final configuration (process $\pname{Use}(y)$) has a state
consisting of a single basis vector in which $y = \ket{1}$. To verify
the protocol for an arbitrary qubit, we can repeat the calculation
with initial state $x,y,z = \frac{\alpha}{\sqrt{2}}(\ket{000}+\ket{110})+\frac{\beta}{\sqrt{2}}(\ket{001}+\ket{111})$.

Alice and Bob are parameterized by their parts ($x,y$) of the
entangled pair (and by the channel $c$). We can be more explicit about
the origin of the entangled pair by introducing what is known in the
physics literature as an
\emph{EPR source}\footnote{EPR stands for
Einstein, Podolsky and Rosen.} (computer scientists might regard it as an
\emph{entanglement server}). This process constructs the entangled pair
(by using the Hadamard and controlled not transformations)
and sends its components to Alice and Bob on the typed channels
$\tid{s,t}{\Chant{\Qbit}}$. Figure~\ref{fig-teleportation-source}
shows the revised model.

\subsection{Bit-Commitment}
\label{sec-bitcommitment}
The bit-commitment problem is to design a protocol such that Alice
chooses a one-bit value which Bob then attempts to guess. The key
issue is that Alice must evaluate Bob's guess with respect to her
original choice of bit, without changing her mind; she must be
committed to her choice. Similarly, Bob must not find out Alice's
choice before making his guess. Bit-commitment turns out to be an important
primitive in cryptographic protocols. Classical bit-commitment schemes
rely on assumptions on the computational complexity of certain
functions; it is natural to ask whether quantum techniques can remove
these assumptions. 

We will discuss a quantum bit-commitment protocol due to Bennett and
Brassard \cite{BennettCH:quacpd} which is closely related to the
quantum key-distribution protocol proposed in the same paper and
known as BB84. The following description of the protocol is
based on Gruska's~\cite{GruskaJ:quac} presentation.

\begin{enumerate}
\item Alice randomly chooses a bit $x$ and a sequence
of bits $\mathit{xs}$. She encodes $\mathit{xs}$ as a sequence of
qubits and sends them to Bob. This encoding uses the standard basis
(representing $0$ by $\ket{0}$ and $1$ by $\ket{1}$) if $x=0$, and the
diagonal basis (representing $0$ by $\ket{+}$ and $1$ by $\ket{-}$) if $x=1$.

\item Upon receiving each qubit, Bob randomly chooses to measure it
with respect to either the standard basis or the diagonal basis. For each
measurement he stores the result and his choice of basis. If the basis
he chose matches Alice's $x$ then the result of the measurement is the
same as the corresponding bit from $\mathit{xs}$; if not, then the
result is $0$ or $1$ with equal probability. After receiving all of
the qubits, Bob tells Alice his guess at the value of $x$.

\item Alice tells Bob whether or not he guessed
correctly. To certify her claim she sends $\mathit{xs}$ to Bob.

\item Bob verifies Alice's claim by looking at the measurements in
which he used the basis corresponding to $x$, and checking that the
results are the same as the corresponding bits from $\mathit{xs}$. He
can also check that the results of the other measurements are
sufficiently random (i.e.\ not significantly correlated with the
corresponding bits from $\mathit{xs}$).
\end{enumerate}

Figure~\ref{fig-bitcommitment} shows our model of this protocol in
CQP. The complexity of the definitions reflects the fact that
we have elaborated much of the computation which is implicit in the
original description.
The definitions use the following features which are not present in
our formalization of CQP, but can easily be added.
\begin{itemize}
\item The type constructor $\kw{List}$ and associated functions and
constructors such
as $\kw{hd}$, $\kw{tl}$, $\kw{length}$, $[\,]$, $@$.
\item Product types ($*$) and functions such as $\kw{fst}$,
$\kw{snd}$.
\item $\kw{if-then-else}$ for expressions and processes.
\item Recursive process definitions.
\end{itemize}
$\pname{Alice}$ is parameterized by $x$ and $\mathit{xs}$; they could
be explicitly chosen at random if desired. $\pname{Bob}$ uses $m$ to
record the results of his measurements, and $n$ (received from
$\pname{Alice}$ initially) as a recursion parameter. $\pname{Bob}$
receives random bits, for his choices of basis, from the server
$\pname{Random}$; he also guesses $x$ randomly. The state
$\pname{BobVerify}$ carries out the first part of step (4) above, but
we have not included a check for non-correlation of the remaining
bits.

Communication between $\pname{Alice}$ and $\pname{Bob}$ uses four
separate channels, $c,\ldots,f$. This proliferation of channels is a
consequence of the fact that our type system associates a unique message
type with each channel. Introducing \emph{session types} \cite{HondaK:intblt}
would allow a single channel to be used for the entire protocol,
although it is worth noting that depending on the physical
implementation of qubits, separation of classical and quantum channels
might be the most accurate model.

We intend to use this CQP model as the basis for various kinds of
formal analysis of the bit-commitment protocol; we make some specific
suggestions in Section~\ref{sec-future}. We should point out, however,
that this bit-commitment protocol is insecure in that it allows Alice
to cheat: if each qubit which she sends to Bob is part of an entangled
pair, then Bob's measurements transmit information back to Alice which
she can use to change $x$ after receiving Bob's guess. The real value
of quantum bit-commitment is as a stepping-stone to the BB84 quantum
key-distribution protocol, which has a very similar structure and is
already being used in practical quantum communication systems.

\begin{figure*}[t]
\begin{gather*}
\tag\Rplus
\vred{\cnfig{\sigma}{\phi}{\plus{u}{v}}}{\cnfig{\sigma}{\phi}{w}}\quad\text{if
$u$ and $v$ are integer literals and $u+v=w$}\\[2mm]
\tag\Rmeasure
\begin{array}{lr}
\multicolumn{2}{l}{\cnfig{q_0,\ldots,q_{n-1}=\alpha_0\ket{\psi_0}+\cdots+\alpha_{2^n-1}\ket{\psi_{2^n-1}}}{\phi}{\msure{q_0,\ldots,q_{r-1}}}\vred{}{}}
\\
\multicolumn{2}{r}{\hspace{15mm}\bp_{0\leqslant
m<2^r}\prob{p_m}{\cnfig{q_0,\ldots,q_n=\frac{\alpha_{l_m}}{p_m}\ket{\psi_{l_m}}+\cdots+\frac{\alpha_{u_m}}{p_m}\ket{\psi_{u_m}}}{\phi}{m}}}
\\
\multicolumn{2}{c}{\text{where $l_m = 2^{n-r}m$, $u_m =
2^{n-r}(m+1)-1$, $p_m = \ms{\alpha_{l_m}}+\cdots+\ms{\alpha_{u_m}}$}}
\end{array}\\[2mm]
\tag\Rtrans
\begin{array}{c}
\vred{\cnfig{q_0,\ldots,q_{n-1} =
\ket{\psi}}{\phi}{\trans{q_0,\ldots,q_{r-1}}{U}}}{\cnfig{q_0,\ldots,q_{n-1}=(U\otimes
I_{n-r})\ket{\psi}}{\phi}{\unit}}
\\
\text{where $U$ is a unitary operator of arity $r$}
\end{array}\\[2mm]
\tag\Rperm
\begin{array}{c}
\vred{\cnfig{q_0,\ldots,q_{n-1} = \ket{\psi}}{\phi}{e}}{\cnfig{q_{\pi(0)},\ldots,q_{\pi(n-1)} = \Pi\ket{\psi}}{\phi}{e}}\\
\text{where $\pi$ is a permutation and $\Pi$ is the corresponding
unitary operator}
\end{array}\\[2mm]
\tag\Rcontext
\frac{\vred{\cnfig{\sigma}{\phi}{e}}{\Prob{i}~\prob{p_{i}}{\cnfig{\sigma_{i}}{\phi_{i}}{e_{i}}}}}{\ered{\cnfig{\sigma}{\phi}{\ctxt{E}{e}}}{\Prob{i}~\prob{p_{i}}{\cnfig{\sigma_{i}}{\phi_{i}}{\ctxt{E}{e_{i}}}}}}
\end{gather*}
\caption{Reduction rules for expression configurations}
\label{fig-reduction-exp}
\end{figure*}

\section{Syntax and Operational Semantics}
\label{sec-syntax}
We now formally define the syntax and operational semantics of the
core of CQP, excluding named process definitions and recursion, which
can easily be added.
\subsection{Syntax}
The syntax of CQP is defined by the grammar in
Figure~\ref{fig-syntax}. Types $T$ consist of data types such as
$\Int$ and $\Unit$ (others can easily be added), the type $\Qbit$ of
qubits, channel types $\Chant{T_1,\ldots,T_n}$ (specifying that each
message is an $n$-tuple with component types $T_1,\ldots,T_n$) and
operator types $\Op{n}$ (the type of a unitary operator on $n$
qubits). The integer range type $\ranget{0}{3}$ used in the
teleportation example is purely for clarification and should be
replaced by $\Int$; we do not expect to typecheck with range types.

We use the notation $\vec{T} = T_1,\ldots,T_n$ and $\vec{e} =
e_1,\ldots,e_n$ and write $\length{\vec{e}}$ for the length of a
tuple. Values $v$ consist of variables ($x$, $y$, $z$ etc.), literal
values of data types ($\num{0},\num{1},\ldots$ and $\unit$) and
unitary operators such as the Hadamard operator
$\qgate{H}$. Expressions $e$ consist of values, measurements
$\msure{e_1,\ldots,e_n}$, applications of unitary operators
$\trans{e_1,\ldots,e_n}{e}$, and expressions involving data operators
such as $e+e'$ (others can easily be added). Note that although the
syntax refers to measurements and transformation of expressions $e$,
the type system will require these expressions to refer to
qubits. Processes $P$ consist of the null (terminated) process $\nil$,
parallel compositions $P\parallel Q$, inputs
$\inp{e}{\tid{\vec{x}}{\vec{T}}}\sep P$ (notation:
$\tid{\vec{x}}{\vec{T}} = \tid{x_1}{T_1},\ldots,\tid{x_n}{T_n}$,
declaring the types of all the input-bound variables), outputs
$\outp{e}{\vec{e}}\sep P$, actions $\action{e}\sep P$ (typically $e$
will be an application of a unitary operator), channel declarations
$(\new\tid{x}{T})P$ and qubit declarations $(\qbit x)P$. In inputs and
outputs, the expression $e$ will be constrained by the type system to
refer to a channel.

The grammar in Figure~\ref{fig-internal-syntax} defines the
\emph{internal} syntax of CQP, which is needed in order to define the
operational semantics. Values are extended by two new forms: qubit
names $q$, and channel names $c$. Evaluation contexts $\ctxt{E}{\,}$
(for expressions) and $\ctxt{F}{\,}$ (for processes) are used in the
definition of the operational semantics, in the style of Wright and
Felleisen \cite{WrightAK:synats}. The structure of $\ctxt{E}{\,}$
is used to define call-by-value evaluation of expressions; the hole
$\ctxt{\,}{\,}$ specifies the first part of the expression to be
evaluated. The structure of $\ctxt{F}{\,}$ is used to define reductions
of processes, specifying which expressions within a process must be evaluated.

Given a process $P$ we define its free variables $\fv{P}$, free qubit
names $\fq{P}$ and free channel names $\fc{P}$ in the usual way;
the binders (of $x$ or $\vec{x}$) are
$\inp{y}{\tid{\vec{x}}{\vec{T}}}$, $(\qbit x)$ and 
$(\new \tid{x}{T})$. 

\subsection{Operational Semantics}

The operational semantics of CQP is defined by reductions (small-step
evaluations of expressions, or inter-process communications)
and probabilistic transitions. The general form of a
reduction is $\cred{t}{\Prob{i}~\prob{p_{i}}{t_{i}}}$ where $t$ and
the $t_i$ are configurations consisting of expressions or processes
with state information. The notation $\Prob{i}~\prob{p_{i}}{t_{i}}$
denotes a probability distribution over configurations, in which
$\Sigma_{i}p_{i} = 1$; we may also write this distribution as 
$\prob{p_{1}}{t_{1}} \bp \cdots \bp \prob{p_{n}}{t_{n}}$. If the
probability distribution contains a single configuration (with
probability $1$) then we simply write $\cred{t}{t'}$. Probability
distributions reduce probabilistically to single configurations:
$\ptrns{\Prob{i}~\prob{p_{i}}{t_{i}}}{p_{i}}{t_{i}}$ (with probability
$p_{i}$, the distribution $\Prob{i}~\prob{p_{i}}{t_{i}}$ reduces to
$t_{i}$).

The semantics of expressions is defined by the reduction relations
$\vred{}{}$ and $\ered{}{}$ (Figure~\ref{fig-reduction-exp}), both on
configurations of the form $\cnfig{\sigma}{\phi}{e}$. If $n$ qubits
have been declared then $\sigma$ has the form $q_0,\ldots,q_{n-1} =
\ket{\psi}$ where $\ket{\psi} =
\alpha_0\ket{\psi_0}+\cdots+\alpha_{2^n-1}\ket{\psi_{2^n-1}}$ is an
element of the $2^n$-dimensional vector space with basis $\ket{\psi_0}
= \ket{0\ldots 0},\ldots,\ket{\psi_{2^n-1}} = \ket{1\ldots 1}$.  The
remaining part of the configuration, $\phi$, is a list of channel
names. Reductions $\vred{}{}$ are basic steps of evaluation, defined
by the rules $\Rplus$ (and similar rules for any other data
operators), $\Rmeasure$ and $\Rtrans$. Rule $\Rperm$ allows qubits in
the state to be permuted, compensating for the way that $\Rmeasure$
and $\Rtrans$ operate on qubits listed first in the state. Measurement
specifically measures the values of a collection of qubits; in the
future we should generalize to measuring \emph{observables} as allowed
by quantum physics.

Reductions
$\ered{}{}$ extend execution to evaluation contexts $\ctxt{E}{\,}$, as
defined by rule $\Rcontext$. Note that the probability distribution
remains at the top level.

Figure~\ref{fig-reduction-proc} defines the reduction relation
$\cred{}{}$ on configurations of the form
$\cnfig{\sigma}{\phi}{P}$. Rule $\Rexpr$ lifts reductions of
expressions to $\ctxt{F}{\,}$ contexts, again keeping probability
distributions at the top level. Rule $\Rcom$ defines communication in
the style of pi-calculus, making use of substitution, which is defined
in the usual way (we assume that bound identifiers are
renamed to avoid capture). Rule $\Ract$ trivially removes actions; in
general the reduction of the action expression to $v$ will have
involved side-effects such as measurement or transformation of quantum
state. Rules $\Rnew$ and $\Rqbit$ create new channels and qubits,
updating the state information in the configuration. Note that this
treatment of channel creation is different from standard presentations
of the pi-calculus; we treat both qubits and channels as elements of a
global store. Rule $\Rpar$ allows reduction to take place in parallel
contexts, again lifting the probability distribution to the top level,
and rule $\Rcong$ allows the use of a structural congruence relation
as in the pi-calculus. Structural congruence is the smallest
congruence relation (closed under the process constructions)
containing $\alpha$-equivalence and closed under the rules in
Figure~\ref{fig-structural-congruence}.

\begin{figure*}
\begin{gather*}
\tag\Rexpr
\frac{\ered{\cnfig{\sigma}{\phi}{e}}{\Prob{i}~\prob{p_{i}}{\cnfig{\sigma_{i}}{\phi_{i}}{e_{i}}}}}{\cred{\cnfig{\sigma}{\phi}{\ctxt{F}{e}}}{\Prob{i}~\prob{p_{i}}{\cnfig{\sigma_{i}}{\phi_{i}}{\ctxt{F}{e_{i}}}}}}
\\[2mm]
\tag\Rcom
\cred{\cnfig{\sigma}{\phi}{\outp{c}{\vec{v}}\sep P \parallel
\inp{c}{\tid{\vec{x}}{\vec{T}}}\sep Q}}{\cnfig{\sigma}{\phi}{P
\parallel \subst{Q}{\vec{v}}{\vec{x}}}}\quad\text{if
$\length{\vec{v}}=\length{\vec{x}}$} \\[2mm]
\tag\Ract
\cred{\cnfig{\sigma}{\phi}{\action{v}\sep P}}{\cnfig{\sigma}{\phi}{P}}
\\[2mm]
\tag\Rnew
\cred{\cnfig{\sigma}{\phi}{(\new\tid{x}{T})P}}{\cnfig{\sigma}{\phi,c}{\subst{P}{c}{x}}}
\quad\text{where $c$ is fresh} \\[2mm]
\tag\Rqbit
\begin{array}{c}
\cred{\cnfig{q_0,\ldots,q_n = \ket{\psi}}{\phi}{(\qbit
x)P}}{\cnfig{q_0,\ldots,q_n,q = \ket{\psi}\otimes\ket{0}}{\phi}{\subst{P}{q}{x}}}
\quad\text{where $q$ is fresh}
\end{array} \\[2mm]
\tag\Rpar
\frac{\cred{\cnfig{\sigma}{\phi}{P}}{\Prob{i}~\prob{p_{i}}{\cnfig{\sigma_{i}}{\phi_{i}}{P_{i}}}}}{\cred{\cnfig{\sigma}{\phi}{P\parallel
Q}}{\Prob{i}~\prob{p_{i}}{\cnfig{\sigma_{i}}{\phi_{i}}{P_i \parallel
Q}}}} \\[2mm]
\tag\Rcong
\frac{P'\scong P \quad
\cred{\cnfig{\sigma}{\phi}{P}}{\Prob{i}~\prob{p_{i}}{\cnfig{\sigma_{i}}{\phi_{i}}{P_{i}}}}\quad
\forall
i.(P_{i}\scong
P_{i}')}{\cred{\cnfig{\sigma}{\phi}{P'}}{\Prob{i}~\prob{p_{i}}{\cnfig{\sigma_{i}}{\phi_{i}}{P_{i}'}}}}
\\[2mm]
\tag\Rprob
\ptrns{\Prob{i}~\prob{p_{i}}{\cnfig{\sigma_i}{\phi_i}{P_{i}}}}{p_{i}}{\cnfig{\sigma_i}{\phi_i}{P_{i}}}
\end{gather*}
\caption{Reduction rules for process configurations}
\label{fig-reduction-proc}
\end{figure*}

\section{Type System}
\label{sec-types}

The typing rules defined in Figure~\ref{fig-typing} apply to the syntax
defined in Figure~\ref{fig-syntax}. Environments $\Gamma$ are mappings
from variables to types in the usual way. Typing judgements are of two kinds.
$\typed{\Gamma}{e}{T}$ means that expression $e$ has type $T$ in
environment $\Gamma$. $\ptyped{\Gamma}{P}$ means that process $P$ is
well-typed in environment $\Gamma$. The rules for expressions are
straightforward; note that in rule $\Ttrans$, $x_1,\ldots,x_n$ must be
distinct variables of type $\Qbit$.

In rule $\Tpar$ the operation $+$ on environments
(Definition~\ref{def-addition-env}) is the key to
ensuring that each qubit is controlled by a unique part of a system. 
An implicit hypothesis of $\Tpar$ is that $\Gamma_1+\Gamma_2$ must be defined.
This is very similar to the linear type system for the pi-calculus,
defined by Kobayashi \emph{et al.} \cite{KobayashiN:linpcfull}.

\begin{definition}[Addition of Environments]\mbox{}\\
The partial operation of adding a typed variable to an environment,
$\Gamma + \tid{x}{T}$, is defined by
\[
\begin{array}{rcll}
\Gamma + \tid{x}{T} & = & \Gamma,\tid{x}{T} & \text{if $x\not\in\dom{\Gamma}$} \\
\Gamma + \tid{x}{T} & = & \Gamma & \text{if $T\not=\Qbit$ and
$\tid{x}{T}\in\Gamma$} \\
\Gamma + \tid{x}{T} & = & \multicolumn{2}{l}{\text{undefined, otherwise}}
\end{array}
\]
This operation is extended inductively to a partial operation
$\Gamma+\Delta$ on environments.
\label{def-addition-env}
\end{definition}

Rule $\Tout$ allows output of classical values and qubits to be
combined, but the qubits must be distinct variables and they cannot be
used by the continuation of the outputting process (note the
hypothesis $\ptyped{\Gamma}{P}$). The remaining rules are
straightforward.

According to the operational semantics, execution of $(\qbit)$ and
$(\new)$ declarations introduces qubit
names and channel names. In order to be able to use the type system to
prove results about the behaviour of executing processes, we introduce
the internal type system (Figure~\ref{fig-typing-int}). This uses
judgements $\ityped{\Gamma}{\Sigma}{\Phi}{e}{T}$ and
$\iptyped{\Gamma}{\Sigma}{\Phi}{P}$ where $\Sigma$ is a set of qubit
names and $\Phi$ is a mapping from channel names to channel
types. Most of the typing rules are straightforward extensions of the
corresponding rules in Figure~\ref{fig-typing}. Because references to
qubits may now be either variables or explicit qubit names, the rules
represent them by general expressions $e$ and impose conditions that
$e$ is either a variable or a qubit name. This is seen in rules
$\ITtrans$ and $\ITout$. Note that in $\ITpar$, the operation
$\Sigma_1+\Sigma_2$ is disjoint union and an implicit hypothesis is
that $\Sigma_1$ and $\Sigma_2$ are disjoint. 

By standard techniques for linear type systems, the typing rules in
Figure~\ref{fig-typing} can be converted into a typechecking algorithm
for CQP models.

As an illustration of the linear control of qubits, consider the
coin-flipping example (Figure~\ref{fig-coinflip}). In $\pname{P}$,
any non-trivial continuation replacing $\nil$ would not be able to use
the qubit $y$, which has been sent on $t$. In $\pname{Q}$, after the
qubit $x$ has been sent on $s$, the continuation cannot use $x$. Of
course, at run-time, the qubit variable $z$ in
$\inp{t}{\tid{z}{\Qbit}}$ is instantiated by $x$, but that is not a
problem because $\pname{P}$ does not use $x$ after sending it. In
$\pname{System}$, $x$ is used as an actual parameter of $\pname{Q}$
and therefore could not also be used as an actual parameter of
$\pname{P}$ (if $\pname{P}$ had a formal parameter of type $\Qbit$).

\begin{figure*}
\[
\renewcommand{\arraystretch}{0.7}
\begin{array}{ccr}
\typed{\Gamma}{v}{\Int}\quad\text{if $v$ is an integer literal}
&
\typed{\Gamma}{\unit}{\Unit}
&
(\Tintlit/\Tunit)
\\ \\
\typed{\Gamma}{\qgate{H}}{\Op{2}}\quad\text{etc.}
&
\typed{\Gamma,\tid{x}{T}}{x}{T}
&
(\Top/\Tvar)
\\ \\
\begin{prooftree}
\typed{\Gamma}{e}{\Int}\quad\typed{\Gamma}{e'}{\Int}
\justifies
\typed{\Gamma}{\plus{e}{e'}}{\Int}
\end{prooftree}
&
\begin{prooftree}
\typed{\Gamma}{\vec{e}}{\vec{\Qbit}}
\justifies
\typed{\Gamma}{\msure{\vec{e}}}{\Int}
\end{prooftree}
&
(\Tplus/\Tmeasure)
\\ \\
\multicolumn{2}{c}{
\begin{prooftree}
\forall i.(\tid{x_i}{\Qbit}\in\Gamma)\quad\text{$x_1\ldots x_n$
distinct}\quad\typed{\Gamma}{U}{\Op{n}}
\justifies
\typed{\Gamma}{\trans{x_1,\ldots,x_n}{U}}{\Unit}
\end{prooftree}
}
&
(\Ttrans)
\\ \\
\ptyped{\Gamma}{\nil}
&
\begin{prooftree}
\ptyped{\Gamma_1}{P}\quad\ptyped{\Gamma_2}{Q}
\justifies
\ptyped{\Gamma_1+\Gamma_2}{P\parallel Q}
\end{prooftree}
&
(\Tnil/\Tpar)
\\ \\
\begin{prooftree}
\typed{\Gamma}{x}{\Chant{T_1,\ldots,T_n}}\quad\ptyped{\Gamma,\tid{y_1}{T_1},\ldots,\tid{y_n}{T_n}}{P}
\justifies
\ptyped{\Gamma}{\inp{x}{\tid{y_1}{T_1},\ldots,\tid{y_n}{T_n}}\sep
P}
\end{prooftree}
&
\begin{prooftree}
\ptyped{\Gamma,\tid{x}{\Qbit}}{P}
\justifies
\ptyped{\Gamma}{(\qbit x)P}
\end{prooftree}
&
(\Tin/\Tqbit)
\\ \\
\multicolumn{2}{c}{
\begin{prooftree}
\typed{\Gamma}{x}{\Chant{T_1,\ldots,T_m,\Qbit,\ldots,\Qbit}}
\quad\forall i.(T_i\not=\Qbit)
\quad\forall i.(\typed{\Gamma}{e_i}{T_i})
\quad\text{$y_i$ distinct} \quad \ptyped{\Gamma}{P}
\justifies
\ptyped{\Gamma,\tid{y_1}{\Qbit}\ldots,\tid{y_n}{\Qbit}}{\outp{x}{e_1,\ldots,e_m,y_1,\ldots,y_n}\sep
P}
\end{prooftree}
}
&
(\Tout)
\\ \\
\begin{prooftree}
\typed{\Gamma}{e}{T}\quad\ptyped{\Gamma}{P}
\justifies
\ptyped{\Gamma}{\action{e}\sep P}
\end{prooftree}
&
\begin{prooftree}
\ptyped{\Gamma,\tid{x}{\Chant{T_1,\ldots,T_n}}}{P}
\justifies
\ptyped{\Gamma}{(\new\tid{x}{\Chant{T_1,\ldots,T_n}})P}
\end{prooftree}
&
(\Taction/\Tnew)
\end{array}
\renewcommand{\arraystretch}{1}
\]
\caption{Typing rules}
\label{fig-typing}
\end{figure*}

\begin{figure*}
\[
\renewcommand{\arraystretch}{0.7}
\begin{array}{ccr}
\ityped{\Gamma}{\Sigma}{\Phi}{v}{\Int}\quad\text{if $v$ is an integer
literal} 
&
\ityped{\Gamma}{\Sigma}{\Phi}{\unit}{\Unit}
&
\hspace{-10pt}(\ITintlit/\ITunit)
\\ \\
\ityped{\Gamma}{\Sigma}{\Phi}{\qgate{H}}{\Op{2}}\quad\text{etc.}
&
\ityped{\Gamma,\tid{x}{T}}{\Sigma}{\Phi}{x}{T}
&
(\ITop/\ITvar)
\\ \\
\ityped{\Gamma}{\Sigma,q}{\Phi}{q}{\Qbit}
&
\ityped{\Gamma}{\Sigma}{\Phi,\tid{c}{T}}{c}{T}
&
(\ITidQ/\ITidC)
\\ \\
\begin{prooftree}
\ityped{\Gamma}{\Sigma}{\Phi}{e}{\Int}\quad\ityped{\Gamma}{\Sigma}{\Phi}{e'}{\Int}
\justifies
\ityped{\Gamma}{\Sigma}{\Phi}{\plus{e}{e'}}{\Int}
\end{prooftree}
&
\begin{prooftree}
\ityped{\Gamma}{\Sigma}{\Phi}{\vec{e}}{\vec{\Qbit}}
\justifies
\ityped{\Gamma}{\Sigma}{\Phi}{\msure{\vec{e}}}{\Int}
\end{prooftree}
&
\hspace{-10pt}(\ITplus/\ITmeasure)
\\ \\
\multicolumn{2}{c}{
\begin{prooftree}
\forall i.(\ityped{\Gamma}{\Sigma}{\Phi}{e_i}{\Qbit}) \quad
\ityped{\Gamma}{\Sigma}{\Phi}{U}{\Op{n}} \quad
\text{each $e_i$ is either $x_i$ or $q_i$, all distinct}
\justifies
\ityped{\Gamma}{\Sigma}{\Phi}{\trans{e_1,\ldots,e_n}{U}}{\Unit}
\end{prooftree}
}
&
(\ITtrans)
\\ \\
\iptyped{\Gamma}{\Sigma}{\Phi}{\nil}
&
\begin{prooftree}
\iptyped{\Gamma_1}{\Sigma_1}{\Phi}{P}\quad\iptyped{\Gamma_2}{\Sigma_2}{\Phi}{Q}
\justifies
\iptyped{\Gamma_1+\Gamma_2}{\Sigma_1+\Sigma_2}{\Phi}{P\parallel Q}
\end{prooftree}
&
(\ITnil/\ITpar)
\\ \\
\begin{prooftree}
\ityped{\Gamma}{\Sigma}{\Phi}{e}{\Chant{T_1,\ldots,T_n}}\quad\iptyped{\Gamma,\tid{y_1}{T_1},\ldots,\tid{y_n}{T_n}}{\Sigma}{\Phi}{P}
\justifies
\iptyped{\Gamma}{\Sigma}{\Phi}{\inp{e}{\tid{y_1}{T_1},\ldots,\tid{y_n}{T_n}}\sep
P}
\end{prooftree}
&
\begin{prooftree}
\iptyped{\Gamma,\tid{x}{\Qbit}}{\Sigma}{\Phi}{P}
\justifies
\iptyped{\Gamma}{\Sigma}{\Phi}{(\qbit x)P}
\end{prooftree}
&
(\ITin/\ITqbit) 
\\ \\
\multicolumn{2}{c}{
\begin{prooftree}
\begin{array}{ccc}
\ityped{\Gamma}{\Sigma}{\Phi}{e}{\Chant{\vec{T},\vec{\Qbit}}} &
\forall i.(T_i\not=\Qbit) &
\forall i.(\ityped{\Gamma}{\Sigma}{\Phi}{e_i}{T_i}) \\
\forall i.(\ityped{\Gamma}{\Sigma}{\Phi}{f_i}{\Qbit}) &
\iptyped{\Gamma}{\Sigma}{\Phi}{P} \\
\multicolumn{3}{c}{\text{$\vec{f}$ consists of distinct variables $\vec{f_x}$
and distinct qubit names $\vec{f_q}$}} 
\end{array}
\justifies
\iptyped{\Gamma,\tid{\vec{f_x}}{\vec{\Qbit}}}{\Sigma,\tid{\vec{f_q}}{\vec{\Qbit}}}{\Phi}{\outp{e}{e_1,\ldots,e_m,f_1,\ldots,f_n}\sep
P}
\end{prooftree}
}
&
(\ITout)
\\ \\
\begin{prooftree}
\ityped{\Gamma}{\Sigma}{\Phi}{e}{T}\quad\iptyped{\Gamma}{\Sigma}{\Phi}{P}
\justifies
\iptyped{\Gamma}{\Sigma}{\Phi}{\action{e}\sep P}
\end{prooftree}
&
\begin{prooftree}
\iptyped{\Gamma,\tid{x}{\Chant{T_1,\ldots,T_n}}}{\Sigma}{\Phi}{P}
\justifies
\iptyped{\Gamma}{\Sigma}{\Phi}{(\new\tid{x}{\Chant{T_1,\ldots,T_n}})P}
\end{prooftree}
&
(\ITaction/\ITnew)
\end{array}
\renewcommand{\arraystretch}{1}
\]
\caption{Internal typing rules}
\label{fig-typing-int}
\end{figure*}

\section{Soundness of the Type System}
\label{sec-soundness}
We prove a series of standard lemmas, following the approach of Wright
and Felleisen \cite{WrightAK:synats}, leading to a proof that typing
is preserved by execution of processes
(Theorem~\ref{theorem-type-preservation}). We then prove that in a
typable process, each qubit is used by at most one of any parallel
collection of sub-processes
(Theorem~\ref{theorem-unique-ownership-qubits}); because of type
preservation, this property holds at every step of the execution of a
typable process. This reflects the physical reality of the protocols
which we want to model.

We can also prove a standard runtime safety theorem, stating that a typable
process generates no communication errors or incorrectly-applied
operators, but we have not included it in the present paper. 
\begin{lemma}[Typability of Subterms in $E$]
\label{lemma-typability-subterms-E}\mbox{}\\
If $\mathcal{D}$ is a typing derivation concluding
$\ityped{\Gamma}{\Sigma}{\Phi}{\ctxt{E}{e}}{T}$ then there exists $U$
such that $\mathcal{D}$ has a subderivation $\mathcal{D}'$ concluding 
$\ityped{\Gamma}{\Sigma}{\Phi}{e}{U}$ and the position of
$\mathcal{D}'$ in $\mathcal{D}$ corresponds to the position of the
hole in $\ctxt{E}{\,}$.
\end{lemma}
\begin{proof}
By induction on the structure of $\ctxt{E}{\,}$.\qed
\end{proof}

\begin{lemma}[Replacement in $E$]
\label{lemma-replacement-E}If
\begin{enumerate}
\item $\mathcal{D}$ is a derivation concluding
$\ityped{\Gamma}{\Sigma}{\Phi}{\ctxt{E}{e}}{T}$
\item $\mathcal{D}'$ is a subderiv.\ of $\mathcal{D}$ concluding 
$\ityped{\Gamma}{\Sigma}{\Phi}{e}{U}$
\item the position of $\mathcal{D}'$ in $\mathcal{D}$ matches the hole in $\ctxt{E}{\,}$
\item $\ityped{\Gamma}{\Sigma}{\Phi}{e'}{U}$
\end{enumerate}
then $\ityped{\Gamma}{\Sigma}{\Phi}{\ctxt{E}{e'}}{T}$.
\end{lemma}
\begin{proof}
Replace $\mathcal{D}'$ in $\mathcal{D}$ by a deriv.\ of
$\ityped{\Gamma}{\Sigma}{\Phi}{e'}{U}$.\qed
\end{proof}

\begin{lemma}[Type Preservation for $\vred{}{}$]
\label{lemma-type-preservation-v}\mbox{}\\
If $\ityped{\Gamma}{\Sigma}{\Phi}{e}{T}$ and
$\vred{\cnfig{\sigma}{\phi}{e}}{\Prob{i} \prob{p_i}{\cnfig{\sigma_i}{\phi_i}{e_i}}}$
and $\Sigma=\dom{\sigma}$ and $\phi=\dom{\Phi}$ then $\forall
i.(\sigma_i=\sigma)$ and $\forall i.(\phi_i=\phi)$ and $\forall i.(\ityped{\Gamma}{\Sigma}{\Phi}{e_i}{T})$.
\end{lemma}
\begin{proof}
Straightforward from the definition of $\vred{}{}$ by examining each
case.\qed
\end{proof}

\begin{lemma}[Type Preservation for $\ered{}{}$]
\label{lemma-type-preservation-e}\mbox{}\\
If $\ityped{\Gamma}{\Sigma}{\Phi}{e}{T}$ and
$\ered{\cnfig{\sigma}{\phi}{e}}{\Prob{i} \prob{p_i}{\cnfig{\sigma_i}{\phi_i}{e_i}}}$
and $\Sigma=\dom{\sigma}$ and $\phi=\dom{\Phi}$ then $\forall
i.(\sigma_i=\sigma)$ and $\forall i.(\phi_i=\phi)$ and $\forall i.(\ityped{\Gamma}{\Sigma}{\Phi}{e_i}{T})$.
\end{lemma}
\begin{proof}
$\ered{\cnfig{\sigma}{\phi}{e}}{\Prob{i}
\prob{p_i}{\cnfig{\sigma_i}{\phi_i}{e_i}}}$ is derived by
$\Rcontext$, so for some $\ctxt{E}{\,}$ we have $e =
\ctxt{E}{f}$ and $\forall i.(e_i = \ctxt{E}{f_i})$ and
$\vred{\cnfig{\sigma}{\phi}{f}}{\Prob{i}
\prob{p_i}{\cnfig{\sigma_i}{\phi_i}{f_i}}}$. From
$\ityped{\Gamma}{\Sigma}{\Phi}{\ctxt{E}{f}}{T}$,
Lemma~\ref{lemma-typability-subterms-E} gives
$\ityped{\Gamma}{\Sigma}{\Phi}{f}{U}$ for some $U$,
Lemma~\ref{lemma-type-preservation-v} gives $\forall
i.(\ityped{\Gamma}{\Sigma}{\Phi}{f-i}{U})$ and $\forall
i.(\sigma_i=\sigma)$ and $\forall i.(\phi_i=\phi)$, and
Lemma~\ref{lemma-replacement-E} gives $\forall i.(\ityped{\Gamma}{\Sigma}{\Phi}{\ctxt{E}{f_i}}{T})$.\qed
\end{proof}

\begin{lemma}[Typability of Subterms in $F$]
\label{lemma-typability-subterms-F}\mbox{}\\
If $\mathcal{D}$ is a typing derivation concluding
$\iptyped{\Gamma}{\Sigma}{\Phi}{\ctxt{F}{e}}$ then there exists $T$
such that $\mathcal{D}$ has a subderivation $\mathcal{D}'$ concluding 
$\ityped{\Gamma}{\Sigma}{\Phi}{e}{T}$ and the position of
$\mathcal{D}'$ in $\mathcal{D}$ corresponds to the position of the
hole in $\ctxt{F}{\,}$.
\end{lemma}
\begin{proof}
By case-analysis on the structure of $\ctxt{F}{\,}$.\qed
\end{proof}

\begin{lemma}[Replacement in $F$]
\label{lemma-replacement-F}If
\begin{enumerate}
\item $\mathcal{D}$ is a derivation concluding
$\iptyped{\Gamma}{\Sigma}{\Phi}{\ctxt{F}{e}}$
\item $\mathcal{D}'$ is a subderiv.\ of $\mathcal{D}$ concluding 
$\ityped{\Gamma}{\Sigma}{\Phi}{e}{T}$
\item the position of $\mathcal{D}'$ in $\mathcal{D}$ matches the hole in $\ctxt{F}{\,}$
\item $\ityped{\Gamma}{\Sigma}{\Phi}{e'}{T}$
\end{enumerate}
then $\iptyped{\Gamma}{\Sigma}{\Phi}{\ctxt{E}{e'}}$.
\end{lemma}
\begin{proof}
Replace $\mathcal{D}'$ in $\mathcal{D}$ by a deriv.\ of
$\ityped{\Gamma}{\Sigma}{\Phi}{e'}{T}$.\qed
\end{proof}

\begin{lemma}[Weakening for Expressions]
\label{lemma-weakening-expressions}\mbox{}\\
If $\ityped{\Gamma}{\Sigma}{\Phi}{e}{T}$ and $\Gamma\subseteq\Gamma'$
and $\Sigma\subseteq\Sigma'$ and $\Phi\subseteq\Phi'$ then
$\ityped{\Gamma'}{\Sigma'}{\Phi'}{e}{T}$.
\end{lemma}
\begin{proof}
Induction on the derivation of
$\ityped{\Gamma}{\Sigma}{\Phi}{e}{T}$.\qed
\end{proof}

\begin{lemma}
\label{lemma-free-names-expressions}\mbox{}\\
If $\ityped{\Gamma}{\Sigma}{\Phi}{e}{T}$ then
$\fv{e}\subseteq\dom{\Gamma}$ and $\fq{e}\subseteq\Sigma$ and
$\fc{e}\subseteq\dom{\Phi}$.
\end{lemma}
\begin{proof}
Induction on the derivation of $\ityped{\Gamma}{\Sigma}{\Phi}{e}{T}$.
\qed
\end{proof}

\begin{lemma}
\label{lemma-free-names-processes}\mbox{}\\
If $\iptyped{\Gamma}{\Sigma}{\Phi}{P}$ then
$\fv{P}\subseteq\dom{\Gamma}$ and $\fq{P}\subseteq\Sigma$ and
$\fc{P}\subseteq\dom{\Phi}$.
\end{lemma}
\begin{proof}
Induction on the derivation of $\iptyped{\Gamma}{\Sigma}{\Phi}{P}$.
\qed
\end{proof}

\begin{lemma}[Substitution in Expressions]
\label{lemma-substitution-expressions}\mbox{}\\
Assume that $\ityped{\Gamma,\tid{\vec{x}}{\vec{T}}}{\Sigma}{\Phi}{e}{T}$
and let $\vec{v}$ be values such that, for each $i$:
\begin{enumerate}
\item if $T_i=\Qbit$ then $v_i$ is a variable or a qubit name
\item if $T_i=\Qbit$ and $v_i=y_i$ (a var) then
$y_i\not\in\Gamma,\tid{\vec{x}}{\vec{T}}$
\item if $T_i=\Qbit$ and $v_i=q_i$ (a qubit name) then
$q_i\not\in\Sigma$
\item if $T_i\not=\Qbit$ then $\ityped{\Gamma}{\Sigma}{\Phi}{v_i}{T_i}$.
\end{enumerate}
Let $\vec{y}$ be the variables of type $\Qbit$ from $\vec{v}$
(corresponding to condition (2)) and assume that they are distinct;
let $\vec{q}$ be the 
qubit names from $\vec{v}$ (corresponding to condition (3)) and assume
that they are distinct.
Then
$\ityped{\Gamma,\tid{\vec{y}}{\vec{\Qbit}}}{\Sigma,\vec{q}}{\Phi}{\subst{e}{\vec{v}}{\vec{x}}}{T}$.
\end{lemma}
\begin{proof}
Induction on the deriv.\ of
$\ityped{\Gamma,\tid{\vec{x}}{\vec{T}}}{\Sigma}{\Phi}{e}{T}$.\qed
\end{proof}

\begin{lemma}[Substitution in Processes]
\label{lemma-substitution-processes}\mbox{}\\
Assume that $\iptyped{\Gamma,\tid{\vec{x}}{\vec{T}}}{\Sigma}{\Phi}{P}$
and let $\vec{v}$ be values such that, for each $i$:
\begin{enumerate}
\item if $T_i=\Qbit$ then $v_i$ is a variable or a qubit name
\item if $T_i=\Qbit$ and $v_i=y_i$ (a var) then
$y_i\not\in\Gamma,\tid{\vec{x}}{\vec{T}}$
\item if $T_i=\Qbit$ and $v_i=q_i$ (a qubit name) then
$q_i\not\in\Sigma$
\item if $T_i\not=\Qbit$ then $\ityped{\Gamma}{\Sigma}{\Phi}{v_i}{T_i}$.
\end{enumerate}
Let $\vec{y}$ be the variables of type $\Qbit$ from $\vec{v}$
(corresponding to condition (2)) and assume that they are distinct;
let $\vec{q}$ be the 
qubit names from $\vec{v}$ (corresponding to condition (3)) and assume
that they are distinct.
Then
$\iptyped{\Gamma,\tid{\vec{y}}{\vec{\Qbit}}}{\Sigma,\vec{q}}{\Phi}{\subst{P}{\vec{v}}{\vec{x}}}$.
\end{lemma}
\begin{proof}
By induction on the derivation of
$\iptyped{\Gamma,\tid{\vec{x}}{\vec{T}}}{\Sigma}{\Phi}{P}$. The key
cases are $\Tpar$ and $\Tout$.

For $\Tpar$ the final step in the typing derivation has the form
\[
\frac{
\iptyped{\Gamma_1}{\Sigma_1}{\Phi}{P}\quad
\iptyped{\Gamma_2}{\Sigma_2}{\Phi}{Q}
}
{
\iptyped{\Gamma,\tid{\vec{x}}{\vec{T}}}{\Sigma}{\Phi}{P\parallel Q}
}
\]
where $\Gamma_1+\Gamma_2=\Gamma,\tid{\vec{x}}{\vec{T}}$ and
$\Sigma_1+\Sigma_2=\Sigma$. Each variable of type $\Qbit$ in
$\Gamma,\tid{\vec{x}}{\vec{T}}$ is in exactly one of $\Gamma_1$ and
$\Gamma_2$. Because the free variables of $P$ and $Q$ are contained in
$\Gamma_1$ and $\Gamma_2$ respectively, substitution into $P\parallel
Q$ splits into disjoint substitutions into $P$ and $Q$. The induction
hypothesis gives typings for $\subst{P}{\vec{v}}{\vec{x}}$ and
$\subst{Q}{\vec{v}}{\vec{x}}$, which combine (by $\Tpar$) to give
$\iptyped{\Gamma,\tid{\vec{y}}{\vec{\Qbit}}}{\Sigma,\vec{q}}{\Phi}{\subst{P\parallel
Q}{\vec{v}}{\vec{x}}}$.
\qed
\end{proof}

\begin{lemma}[Struct.\ Cong.\ Preserves Typing]
\label{lemma-structural-congruence-typing}\mbox{}\\
If $\iptyped{\Gamma}{\Sigma}{\Phi}{P}$ and $P\scong Q$ then
$\iptyped{\Gamma}{\Sigma}{\Phi}{Q}$.
\end{lemma}
\begin{proof}
Induction on the derivation of $P\scong Q$.\qed
\end{proof}

\begin{lemma}[External/Internal Type System]\mbox{}\\
$\typed{\Gamma}{e}{T} \Rightarrow
\ityped{\Gamma}{\emptyset}{\emptyset}{e}{T}$ and $\ptyped{\Gamma}{P}
\Rightarrow \iptyped{\Gamma}{\emptyset}{\emptyset}{P}$.
\end{lemma}
\begin{proof}
Induction on the derivations.\qed
\end{proof}

\begin{theorem}[Type Preservation for $\cred{}{}$]
\label{theorem-type-preservation}\mbox{}\\
If $\iptyped{\Gamma}{\Sigma}{\Phi}{P}$ and
$\cred{\cnfig{\sigma}{\phi}{P}}{\Prob{i} \prob{p_i}{\cnfig{\sigma_i}{\phi_i}{P_i}}}$
and $\Sigma=\dom{\sigma}$ and $\phi=\dom{\Phi}$ then $\forall
i.(\sigma_i=\sigma)$ and $\forall i.(\phi_i=\phi)$ and $\forall i.(\iptyped{\Gamma}{\Sigma}{\Phi}{P_i})$.
\end{theorem}
\begin{proof}
By induction on the derivation of
$\cred{\cnfig{\sigma}{\phi}{P}}{\Prob{i}
\prob{p_i}{\cnfig{\sigma_i}{\phi_i}{P_i}}}$, in each case examining
the final steps in the derivation of $\iptyped{\Gamma}{\Sigma}{\Phi}{P}$.\qed
\end{proof}

\begin{theorem}[Unique Ownership of Qubits]
\label{theorem-unique-ownership-qubits}\mbox{}\\
If $\iptyped{\Gamma}{\Sigma}{\Phi}{P\parallel Q}$ then
$\fq{P}\cap\fq{Q}=\emptyset$.
\end{theorem}
\begin{proof}
The final step in the derivation of
$\iptyped{\Gamma}{\Sigma}{\Phi}{P\parallel Q}$ has the form
\[
\frac{
\iptyped{\Gamma_1}{\Sigma_1}{\Phi}{P}
\quad
\iptyped{\Gamma_2}{\Sigma_2}{\Phi}{Q}
}
{
\iptyped{\Gamma}{\Sigma}{\Phi}{P\parallel Q}
}
\]
where $\Gamma = \Gamma_1+\Gamma_2$ and $\Sigma =
\Sigma_1+\Sigma_2$. By Lemma~\ref{lemma-free-names-processes},
$\fq{P}\subseteq\Sigma_1$ and $\fq{Q}\subseteq\Sigma_2$. The implicit
hypothesis of the typing rule $\Tpar$ is that $\Sigma_1+\Sigma_2$ is
defined, meaning that $\Sigma_1\cap\Sigma_2=\emptyset$. Hence
$\fq{P}\cap\fq{Q}=\emptyset$.\qed
\end{proof}

\section{Future Work}
\label{sec-future}
Our aim is to develop techniques for formal verification of systems
modelled in CQP. In particular we are working towards an analysis of
the BB84 quantum key distribution protocol, including both the core
quantum steps and the classical authentication phase. Initially we
will use model-checking, in both standard (non-deterministic) and
probabilistic forms. Standard model-checking is appropriate for
absolute properties (for example, the quantum teleportation protocol
(Section~\ref{sec-teleportation}) claims that the final state of $y$
is always the same as the initial state of $z$). In general, however,
probabilistic model-checking is needed. For
example, the bit-commitment protocol (Section~\ref{sec-bitcommitment})
guarantees that, with some high probability which is
dependent on the number of bits used by Alice, Bob's verification step
is successful. We have obtained preliminary results
\cite{NagarajanR:forvqp,PapanikolaouN:msc} with the CWB-NC
\cite{CleavelandR:cwbnc} and PRISM \cite{KwiatkowskaMZ:pripsm}
systems, working directly with the modelling language of each
tool. The next step is to develop automated translations of CQP into
these lower-level modelling languages; note that our operational
semantics matches the semantic model used by PRISM. 

Another major area for future work is to develop a theory of
equivalence for CQP processes, as a foundation for compositional
techniques for reasoning about the behaviour of systems.

We can also consider extending the language. It should be
straightforward to add purely classical features such as functions and
assignable variables. Extensions which combine quantum data with
enhanced classical control structures require more care. Valiron's
\cite{ValironB:quat} recent formulation of a typed quantum lambda
calculus seems very compatible with our approach, and it should fit
into CQP's expression language fairly easily.

\section{Conclusions}
\label{sec-conclusions}
We have defined a language, CQP, for modelling systems which combine
quantum and classical communication and computation. CQP has a formal
operational semantics, and a static type system which guarantees that
transmitting a qubit on a communication channel corresponds to a
physical transfer of ownership.

The syntax and semantics of CQP are based on a combination of the
pi-calculus and an expression language which includes measurement and
transformation of quantum state. The style of our definitions makes it
easy to enrich the language.

Our research programme is to use CQP as the basis for 
analysis and verification of quantum protocols, and we have outlined
some possibilities for the use of both standard and probabilistic
model-checking.


\end{document}